\documentclass[aps,prb,twocolumn,superscriptaddress]{revtex4-2}
\usepackage{amsfonts}
\usepackage{graphicx}
\usepackage{epstopdf}
\usepackage{epsfig}
\usepackage{amsmath}
\usepackage[normalem]{ulem}
\usepackage{multirow}
\usepackage{color}
\usepackage{makecell}
\usepackage{array}
\usepackage{booktabs}
\usepackage{mathtools}
\usepackage{bm}
\usepackage{color}
\usepackage{array}
\usepackage{booktabs}
\usepackage{tabularx}

\usepackage{natbib}
\usepackage{notoccite}

\newcolumntype{Y}{>{\centering\arraybackslash}X}

\newcommand{\Rmnum}[1]{\expandafter\@slowromancap\romannumeral #1@}
\makeatother

\begin{document}

\title{Optical response of two-dimensional Dirac materials with a flat band}

\author{Chen-Di Han}
\affiliation{School of Electrical, Computer and Energy Engineering, Arizona State University, Tempe, Arizona 85287, USA}

\author{Ying-Cheng Lai} \email{Ying-Cheng.Lai@asu.edu}
\affiliation{School of Electrical, Computer and Energy Engineering, Arizona State University, Tempe, Arizona 85287, USA}
\affiliation{Department of Physics, Arizona State University, Tempe, Arizona 85287, USA}

\date{\today}

\begin{abstract}

Two-dimensional Dirac materials with a flat band have been demonstrated to possess a plethora of unusual electronic properties, but the optical properties of these materials are less studied. Utilizing $\alpha$-$\mathcal{T}_3$ lattice as a prototypical system, where $0\le \alpha \le 1$ is a tunable parameter and a flat band through the conic intersection of two Dirac cones arises for $\alpha > 0$, we investigate the conductivity of flat-band Dirac material systems analytically and numerically. 
Motivated by the fact that the imaginary part of the optical conductivity can have significant effects on the optical response and is an important factor of consideration for developing $\alpha$-$\mathcal{T}_3$ lattice based optical devices, we are led to derive a complete conductivity formula with both the real and imaginary parts. Scrutinizing the formula, we uncover two phenomena. First, for the value of $\alpha$ in some range, two types of optical transitions coexist: one between the two Dirac cones and another from the flat band to a cone, which generate multi-frequency transverse electrical propagating waves. Second, for $\alpha=1$ so the quasiparticles become pseudospin-1, the flat-to-cone transition can result in resonant scattering. These results pave the way to exploiting $\alpha$-$T_3$ lattice for optical device applications in the terahertz frequency domain.

\end{abstract}

\date{\today}

\maketitle

\section{Introduction} \label{sec:intro}

Quantum materials whose energy band consists of a pair of Dirac cones and a
topologically flat band, electronic or optical, constitute a frontier area
of research~\cite{Sutherland1986,bercioux2009massless,Shen:2010,Green2010,Dora2011,wang2011nearly,Huang2011,Mei:2012,Moitra2013,raoux2014dia,Guzman:2014,romhanyi2015hall,giovannetti2015kekule,Li2015,Mu2015,Vic2015,Taiee2015,Fang2016,Diebel:2016,Zhu2016,Bradlynaaf2016,Fulga2017,Ezawa2017,Zhong2017,Zhu2017,Drost2017,slot2017experimental,Tan2018}. 
For example, in a dielectric photonic crystal, Dirac cones can be induced by 
accidental degeneracy occurring at the center of the Brillouin zone, which 
effectively makes the crystal a zero-refractive-index metamaterial at the 
Dirac point where the Dirac cones intersect with another flat
band~\cite{Huang2011,Mei:2012,Moitra2013,Li2015,Fang2016}. Alternatively,
configuring an array of evanescently coupled optical waveguides into a Lieb
lattice~\cite{Guzman:2014,Mu2015,Vic2015,Diebel:2016} can lead to a gapless
spectrum consisting of a pair of common Dirac cones and a perfect flat middle
band at the corner of the Brillouin zone. Loading cold atoms into an optical
Lieb lattice provides another experimental realization of the gapless
three-band spectrum at a smaller scale with greater dynamical controllability
of the system parameters~\cite{Taiee2015}. Dice or $\mathcal{T}_3$ optical
lattices also possess the Dirac cone and flat-band structure~\cite{RCF:2006,Burkov2006,bercioux2009massless,Dora2011,raoux2014dia,Andrijauskas2015}.
Electronically, Dirac materials that can generate a topologically flat band
include transition-metal oxide SrTiO$_3$/SrIrO$_3$/SrTiO$_3$ trilayer
heterostructures~\cite{wang2011nearly}, $2$D carbon or MoS$_2$ allotropes with 
a square symmetry~\cite{Li2014}, as well as 
SrCu$_2$(BO$_3$)$_2$~\cite{romhanyi2015hall} and graphene-In$_2$Te$_2$
bilayer~\cite{giovannetti2015kekule}.

In two-dimensional (2D) Dirac materials with a flat band, the quasiparticles
are of the massless or massive pseudospin-1 type. Comparing with the
conventional Dirac cone systems with massless pseudospin-1/2
quasiparticles (e.g., graphene)~\cite{novoselov2005two,ZTSK:2005},
pseudospin-$1$ systems can exhibit quite unusual and unconventional
physics such as super-Klein tunneling for the two conical, linearly
dispersive bands~\cite{Shen:2010,Urban2011,Dora2011,Fang2016,XL:2016},
diffraction-free wave propagation and novel conical
diffraction~\cite{Guzman:2014,Mu2015,Vic2015,Diebel:2016}, flat-band rendered
divergent DC conductivity with a tunable short-range 
disorder~\cite{vigh2013diverging}, unconventional Anderson 
localization~\cite{Chalker2010,Body2014}, flat-band
ferromagnetism~\cite{Lieb1989,Tasaki1992,Taiee2015}, and peculiar topological
phases under external gauge fields or spin-orbit
coupling~\cite{Aoki1996,Weeks2010,wang2011nearly,Goldman2011}. In particular,
topological phases arise due to the flat band that permits a number of
degenerate localized states with a topological origin, i.e., ``caging'' of
carriers~\cite{Vidal1998}. Additional phenomena include superscattering of
pseudospin-1 waves in the subwavelength regime~\cite{XL:2017}, geometric
valley Hall effect and and valley filtering~\cite{XHHL:2017}, chaos based
Berry phase detection~\cite{WHXL:2019}, atomic collapse in pseudospin-1
systems~\cite{han2019atomic}, anomalous chiral edge states in spin-1 Dirac-Weyl
quantum dots~\cite{XL:2020a}, anomalous in-gap edge states in
two-dimensional pseudospin-1 Dirac-Weyl insulators~\cite{XL:2020b},
and the analogy between Klein scattering of spin-1 Dirac-Weyl wave and
localized surface plasmon~\cite{XHL:2021}. There was also a study of interplay
between classical chaos and flat-band  physics~\cite{HXL:2020}.
In the past few years, magic-angle twisted bilayer graphene, a type of quantum
materials hosting a flat band, has become a forefront area of research. These
materials can generate surprising physical phenomena such as
superconductivity~\cite{Caoetal:2018,YCPZWTGYD:2019}, orbital
ferromagnetism~\cite{Sharpeetal:2019,Luetal:2019}, and the Chern insulating
behavior with topological edge states. Notwithstanding the diversity and the
broad scales to realize the band structure that consists of two conical bands
and a characteristic flat band intersecting at a single point in different
physical systems, theoretically a unified framework exists: the generalized
Dirac-Weyl equation for massless or massive spin-1
particles~\cite{bercioux2009massless,XL:2016}.

The optical properties of 2D Dirac materials without a flat
band have been extensively investigated, such as the optical responses of
graphene~\cite{vakil2011transformation,grigorenko2012graphene,bao2012graphene,
YHL:2018}, its theoretically predicted frequency-dependent optical
conductivity~\cite{gusynin2006transport,falkovsky2007space,gusynin2006unusual,abergel2007optical,mikhailov2007new,hanson2008dyadic} and experimental
verification~\cite{li2008dirac,kuzmenko2008universal,mak2008measurement},
suggesting the possibility of developing graphene-based tunable terahertz
optical devices. Such devices can have applications ranging from light
transform~\cite{vakil2011transformation,bao2011broadband,li2014ultrafast} and
high frequency communication~\cite{ju2011graphene,nikitin2011edge,
akyildiz2014terahertz} to cloaking or
superscattering~\cite{chen2011atomically,li2015tunable}. In particular, the
discovery of novel plasmon mode in graphene led to the first experimental
superscattering system at a macroscopic scale~\cite{qian2019experimental}.
These efforts have given birth to new fields such as topological
photonics~\cite{lu2014topological, ozawa2019topological} and topological
lasing~\cite{harari2018topological, bandres2018topological}. Besides
single-layer graphene, the optical properties of other 2D material have
also been studied. For example, surface plasmon has been uncovered and
characterized in hexagonal boron nitride (hBN) or graphene hBN hybrid
structures~\cite{woessner2015highly,musa2017confined}, in
bilayer~\cite{abergel2007optical,jablan2011transverse} and
multilayer~\cite{nilsson2006electronic,falkovsky2007optical} graphene, in
twisted graphene bilayer~\cite{stauber2013optical,stauber2016quasi,
stauber2018chiral,lin2020chiral,deng2020strong}, in
silicene~\cite{ukhtary2016broadband} and hyperbolic
materials~\cite{iorsh2013hyperbolic,qian2018multifrequency}.
Particularly worth mentioning is the work on the conductivity for gapped Dirac 
fermions~\cite{li2013conductivity}, where the Kubo formula was used to calculate
the contribution of the intra-band transitions to the optical conductivity. A 
central goal in these studies is to garner a strong optical response at certain
frequency or at multiple frequencies. In 2D Dirac materials without a flat band,
transverse magnetic (TM, or $p$-polarized) and transverse electric (TE, or 
$s$-polarized) polaritons have been found to emerge at different frequencies, 
where the TE polarization can arise at high frequencies with a low 
loss~\cite{mikhailov2007new}.


For Dirac materials with a flat band, most previous work focused on their
electronic properties: their optical properties have been less studied. In
this paper, we investigate the ``complete'' optical responses of 2D Dirac
materials with a flat band, in the sense that both the real and imaginary parts
of the optical conductivity are derived, using the $\alpha$-$\mathcal{T}_3$
lattice as a paradigmatic model system of such materials. This lattice is
formed by adding an atom at the center of each unit cell of the honeycomb
graphene lattice~\cite{bercioux2009massless}, where the low energy excitations
can be described by the pseudospin-1 Dirac-Weyl equation. The parameter
$\alpha$ characterizes the interaction strength between the central atom and
any of its nearest neighbors, relative to that between two neighboring atoms
at the vertices of the hexagonal cell. For $\alpha=0$ there is no coupling
between the central atom and a vertex , so the lattice degenerates to graphene
with pseudospin-1/2 quasiparticles. As the value of $\alpha$ increases from
zero to one, a flat band through the conic interaction of the two Dirac cones
emerges and its physical influences become progressively
pronounced~\cite{raoux2014dia,illes2015hall}. For $\alpha=1$, the lattice 
generates pseudospin-1 quasiparticles. The flat band can lead to physical 
phenomena such as the divergence of
conductivity~\cite{vigh2013diverging,louvet2015origin,han2019atomic}.
Under a continuous approximation, an $\alpha$-$\mathcal{T}_3$ lattice can be
treated as a thin layer with certain surface conductivity. Unlike graphene,
here the surface conductivity is contributed to by three types of transitions
between the bands: intraband transition, cone-to-cone transition, and
flat-band-to-cone transition. Previously, the optical conductivity of
$\alpha$-$\mathcal{T}_3$ lattice was ``partially'' studied in the sense
that only the real part of the conductivity has been
derived~\cite{illes2015hall,tabert2016optical,kovacs2017frequency,
chen2019enhanced}. Considering that the imaginary part can affect the
optical response significantly and is therefore important for developing
$\alpha$-$\mathcal{T}_3$ lattice based optical devices, we are led to derive
a complete conductivity formula with both the real and imaginary parts. The
formula is verified through two independent approaches and leads to two
previously uncovered phenomena. First, while the intraband transition leads
to TM polarized waves at low frequencies (1-10 THz), TE polarized waves can
emerge at high frequencies (100-300 THz), due to the two interband transitions.
Second, the unique flat-band-to-cone transition generates multi-frequency TE
propagating waves for $\alpha\in(0.4, 0.6)$ and a strong optical response for
$\alpha=1$. These phenomena are confirmed through studying the behaviors of
propagating surface wave and scattering.

We remark that a viable experimental way to realize $\alpha$-$\mathcal{T}_3$
lattice is through photonic
crystals~\cite{RCF:2006,slot2017experimental,LF:2018,MDOTG:2018}, where
the three nonequivalent atoms in a unit cell can be simulated by using
coupled waveguides generated by laser inscription~\cite{MDOTG:2018}.
Electronic materials can also be exploited to generate pseudospin-1 lattice
systems such as transition-metal oxide
$\text{SrTiO}_3/\text{SrIrO}_3/\text{SrTiO}_3$ trilayer
heterostructures~\cite{wang2011nearly},
$\text{SrCu}_2(\text{BO}_3)_2$~\cite{romhanyi2015hall} or
graphene-$\text{In}_2\text{Te}_2$~\cite{giovannetti2015kekule}. Recently,
materials with a flat band  have been reported~\cite{franchina2020engineering}
A review of the experimentally realized dice-like system can be found in
Ref.~\cite{leykam2018artificial}. The physics of certain solid-state materials
is effectively that of an $\alpha$-$\mathcal{T}_3$ lattice. For example, a
theoretical analysis and computations showed~\cite{malcolm2015magneto} that
the material $\text{Hg}_{1-x}\text{Cd}_{x}\text{Te}$ is equivalent to the
$\alpha$-$\mathcal{T}_3$ lattice with $\alpha=1/\sqrt{3}\approx0.58$. This
material has been realized in experiments~\cite{teppe2016temperature,charnukha2019ultrafast,hubmann2020symmetry}. For this reason, we emphasize the case of
$\alpha=1/\sqrt{3}\approx0.58$ in this paper.

In Sec.~\ref{sec:conductivity}, we describe the optical conductivity for
$\alpha$-$\mathcal{T}_3$ lattice, which consists of three parts, and we use
two different methods to derive the conductivity. In Sec.~\ref{sec:plane},
we solve the Maxwell's equations for $\alpha$-$\mathcal{T}_3$ lattice in a
dielectric medium and characterize the properties of the TM and TE polarized
waves using the loss and attenuation length. In Sec.~\ref{sec:sphere}, we
study optical scattering from a sphere coated with $\alpha$-$\mathcal{T}_3$
lattice and discuss potential optical device applications. Conclusions and
a discussion are offered in Sec.~\ref{sec:discussion}.

\section{Optical conductivity of $\alpha$-$\mathcal{T}_3$ lattice} \label{sec:conductivity}

\begin{figure}
\centering
\includegraphics[width=\linewidth]{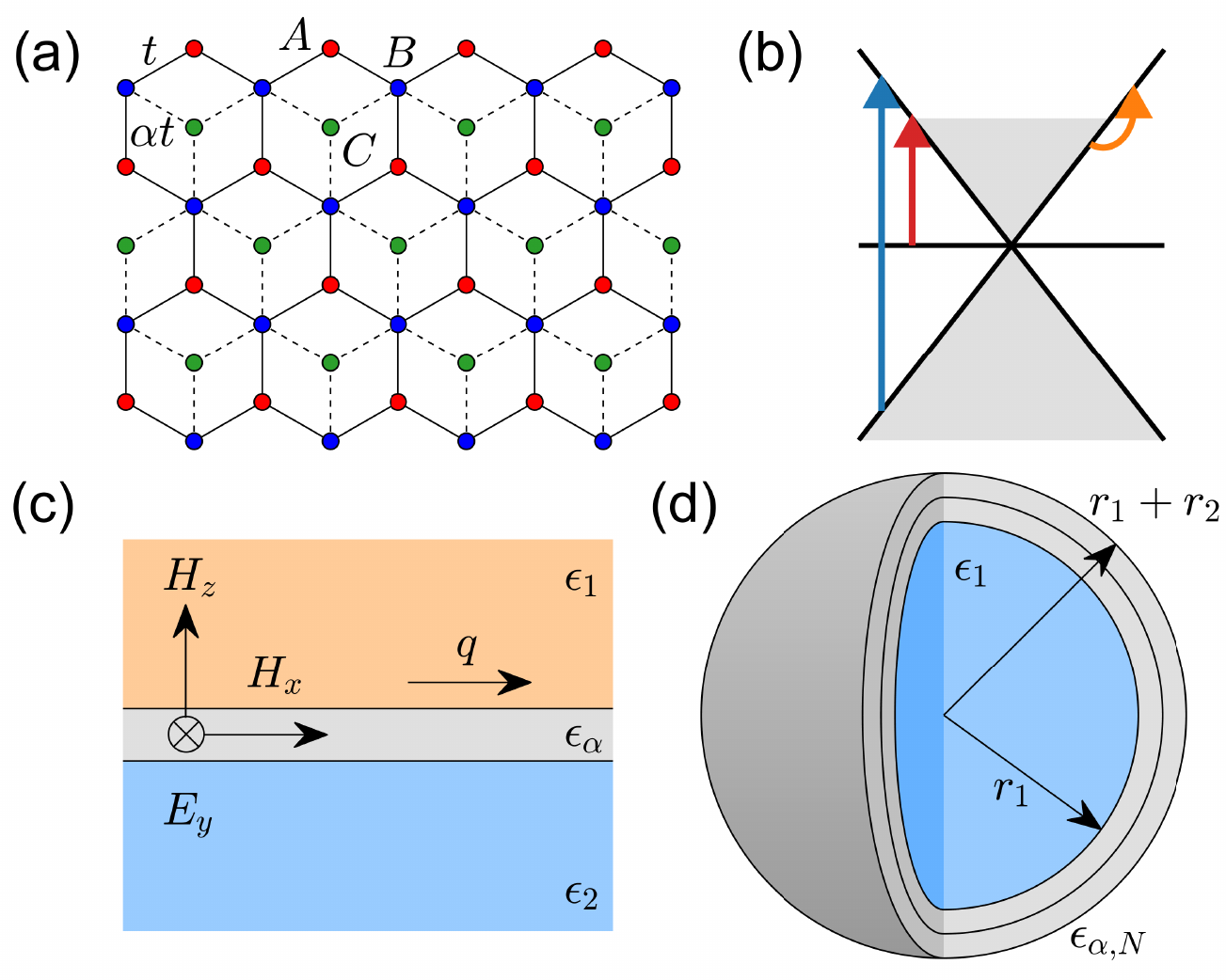}
\caption{Illustration of an $\alpha$-$\mathcal{T}_3$ lattice, its band
structure, optical transition, and a possible $\alpha$-$\mathcal{T}_3$ lattice
based optical devices. (a) An $\alpha$-$\mathcal{T}_3$ lattice, where A, B and
C are the three non-equivalent atoms. The hopping energy between A and B is
$t$ (solid line) and that between B and C is $\alpha t$ (dashed line). (b) Band
transitions in the $\alpha$-$\mathcal{T}_3$ lattice, where the conduction,
valence and flat bands are shown. For a positive chemical potential at zero
temperature, states below it are filled and states above are empty. The three
arrows indicate three band transitions, each contributing to the optical
conductivity. (c) TE mode in an $\alpha$-$\mathcal{T}_3$ lattice placed
between two infinite media with dielectric constants $\epsilon_1$ and
$\epsilon_2$, respectively. (d) A dielectric sphere coated with multilayer
$\alpha$-$\mathcal{T}_3$ lattices. The sphere has dielectric constant
$\epsilon_1$ and radius $r_1$. The multilayer system has depth $r_2$ and $N$
layers of $\alpha$-$\mathcal{T}_3$ lattice with the dielectric function
$\epsilon_{\alpha,N}$. }
\label{fig:schematic}
\end{figure}

An $\alpha$-$\mathcal{T}_3$ lattice is generated by placing an additional atom
at the center of each unit cell of the honeycomb lattice, where there are three
nonequivalent atoms, as shown in Fig.~\ref{fig:schematic}(a). The range of
the variation of the parameter $\alpha$ is $[0,1]$, where the quasiparticles
are pseudospin-1/2 for $\alpha = 0$ and pseudospin-1 for $\alpha = 1$, and
a flat band arises for $0<\alpha \le 1$. For convenience, we say that the
quasiparticles for $0<\alpha<1$ are of the ``hybrid'' type. Under the
tight-binding approximation, the low-energy excitation Hamiltonian is given
by~\cite{raoux2014dia,illes2015hall}
\begin{equation} \label{eq:2_Hamiltonian}
H_\alpha=v_F\hbar\begin{pmatrix}
0 & f_\mathbf{k}\cos(\phi) & 0 \\
f_\mathbf{k}^* \cos(\phi) & 0 & f_\mathbf{k} \sin(\phi)\\
0 & f_\mathbf{k}^* \sin(\phi) & 0
\end{pmatrix},
\end{equation}
where $f_\mathbf{k}=vk_x-ik_y$, $\alpha\equiv\tan(\phi)$, $v_F$ is the Fermi
velocity, and $v=\pm 1$ denotes the valley index. Solving the eigenvalue
problem associated with the tight-binding Hamiltonian matrix, we obtain three
eigenfunctions, as detailed in Appendix~\ref{Appendix_A}. There are three
distinct energy bands: a pair of Dirac cones and a flat band, as shown in
Fig.~\ref{fig:schematic}(b).

The optical properties of the $\alpha$-$\mathcal{T}_3$ lattice are largely
controlled by the surface conductivity $\sigma$, which is determined by the
expectation value of the current operator. In the $x$ direction as specified
in Fig.~\ref{fig:schematic}(c), the current operator is $j_x=-ev_FS_x$, where
\begin{equation} \label{eq:2_Sx}
S_x=\begin{pmatrix}
0 & \cos \phi & 0\\
\cos\phi & 0 & \sin \phi \\
0 & \sin \phi & 0
\end{pmatrix}.
\end{equation}
Expanding the current vector in the three bands, we obtain the matrix
representation of $j_x$, as described in Appendix~\ref{Appendix_A}. For weak
electrical field, the conductivity is given by the standard Kubo
formula~\cite{falkovsky2007space,hanson2008dyadic}
\begin{equation} \label{eq:2_Kubo}
\begin{split}
\sigma_{xx}(\omega,\phi)=\frac{\hbar}{2i\pi^2}\sum_{n,m}\frac{F(E_m)-F(E_n)}{E_n-E_m}\times \\
\left(\frac{\langle n | j_{x}|m\rangle\langle m| j_x|n\rangle}{E_n-E_m-\hbar\omega}+\frac{\langle m | j_x|n\rangle\langle n | j_x|m\rangle}{E_m-E_n-\hbar\omega}\right),
\end{split}
\end{equation}
where the subscript in $\sigma$ indicates the direction of the current and
electric field, and $\omega$ is the frequency of the electromagnetic wave.
For a homogeneous material without a magnetic field, we have
$\sigma_{xx}=\sigma_{yy}$ and $\sigma_{xy}=\sigma_{yx}=0$. The summation is
over all states $|n\rangle=|\mathbf{k}; \pm, 0\rangle$ and
$|m\rangle=|\mathbf{k}'; \pm', 0'\rangle$ with the respective energy $E_n$ and
$E_m$. Evaluating the integral, the nonzero terms appear only
for~\cite{voon1993tight} $\mathbf{k}=\mathbf{k}'$. The quantity $F$
in Eq.~\eqref{eq:2_Kubo} stands for the Fermi-Dirac distribution. At zero
temperature, the only transitions allowed are those from the filled to the
unfilled bands, or vice versa.

To obtain the optical conductivity, it is necessary to evaluate the summation
in the Kubo formula Eq.~\eqref{eq:2_Kubo}. The details of the derivation are
presented in Appendix~\ref{Appendix_B1}. The validity of the derivation and
the result can be established by using the Kramers–Kronig (KK) relation as
described in Appendix~\ref{Appendix_B2}, which gives the same results. Here
we summarize our complete conductivity formulas.

Physically, momentum conservation stipulates that the summation for
$|n\rangle$ and $|m\rangle$ can be regarded as corresponding to band
transition processes. In particular, the summation can be divided into three
parts, denoted as $\sigma^{(1)}(\omega,\phi)$, $\sigma^{(2)}(\omega,\phi)$
and $\sigma^{(3)}(\omega,\phi)$, which correspond to the intraband,
cone-to-cone and flat-band-to-cone transitions, respectively. The conductivity
due to the intraband transition is
\begin{equation} \label{eq:2_Intra_final}
\sigma^{(1)}(\omega,\phi)=4\mu\sigma_0\delta(\hbar\omega)+\frac{4i\mu\sigma_0}{\pi\hbar\omega},
\end{equation}
where $\sigma_0=e^2/(4\hbar)$ and $\mu$ is the chemical potential.
The Drude peak is represented by a $\delta$-function with the coefficient
proportional to the chemical potential $\mu$.
The conductivity due to the cone-to-cone transition is
\begin{equation} \label{eq:2_Inter1_final}
\begin{split}
\sigma^{(2)}(\omega,\phi)=&\cos^2(2\phi)\sigma_0 \times \\
&\left[\Theta(\hbar\omega-2\mu)-\frac{i}{\pi}\ln \left|\frac{\hbar\omega+2\mu}{\hbar\omega-2\mu} \right| \right],
\end{split}
\end{equation}
where $\Theta$ is the Heaviside step function. The conductivity due to the
flat-band-to-cone transition is
\begin{equation} \label{eq:2_Inter2_final}
\begin{split}
\sigma^{(3)}(\omega,\phi)=&2\sin^2(2\phi)\sigma_0\times \\
&\left[\Theta(\hbar\omega-\mu)-\frac{i}{\pi}\ln \left|\frac{\hbar\omega+\mu}{\hbar\omega-\mu} \right| \right].
\end{split}
\end{equation}
For convenience, we introduce a unit free conductivity and divide it into real
and imaginary parts
\begin{equation}
\sigma=\sigma'+i\sigma''=\sigma_0(\tilde{\sigma}'+i\tilde{\sigma}'').
\end{equation}
At finite temperatures, the Fermi-Dirac distribution is no longer a step
function and this will lead to a change in the conductivity formula, as
detailed in Appendix~\ref{Appendix_B3}.

We have also analyzed the effect of finite impurity scattering on the optical
conductivity, as detailed in Appendix~\ref{Appendix_B4}. The main result is 
that finite impurity scattering will change the conductivity as in 
Eq.~\eqref{eq:2_Intra_final} for $\omega\rightarrow 0$.

For 2D Dirac materials such as graphene, the high frequency regime above
$10$THz is physically important for the optical conductivity. Nonetheless,
the behavior of the conductivity in the low frequency regime can be analyzed
in terms of the product $\omega\sigma''$ between the frequency and the 
imaginary part of the conductivity (see Appendix~\ref{Appendix_B5}). This is 
motivated by the fact that, in the study of optical conductivity of 
superconducting materials, the quantity $\omega\sigma''$ is often used to 
characterize the penetration 
depth~\cite{marsiglio1996imaginary,jiang1996imaginary, pronin2001optical}. 
We also study the effects of varying $\alpha$
on the optical conductivity and find that the flat band does not play a
significant role in the conductivity in the low frequency regime. However, 
decreasing the Fermi energy or increasing the temperature can make the 
flat-band contribution more pronounced. The different relaxation time for 
different values of $\alpha$ can also impact the optical conductivity.

For $\alpha=0$ (or equivalently, $\phi=0$), our conductivity formula reduces
to the one for graphene~\cite{gusynin2006transport,falkovsky2007space,gusynin2006unusual,abergel2007optical,mikhailov2007new,hanson2008dyadic},
which was experimentally
verified~\cite{li2008dirac,kuzmenko2008universal, mak2008measurement}.
For $\alpha\ne 0$, the real part of the conductivity is consistent with the
previous result~\cite{illes2015hall}. The imaginary part of the conductivity
is the contribution of our present work.

Features of the derived conductivity formulas
Eqs.~(\ref{eq:2_Intra_final}-\ref{eq:2_Inter2_final}), are as follows.
First, the conductivity does not depend on the detailed lattice
structure, insofar as there are a pair of Dirac cones and a flat band. In
fact, the conductivity is determined by the linear dispersion relationship
and the flat band. Our derivation thus holds for different types of lattices
described by the effective Hamiltonian in \eqref{eq:2_Hamiltonian}. Second,
the formulas hold only for a reasonable range in $\omega$ or $\mu$. If the
value of $\omega$ is small, impurity scattering will be important and, in this
case, it is necessary to make the change $\omega\rightarrow\omega+i\tau^{-1}$,
where $\tau$ is the relaxation time. This change will smooth out the
$\delta$-function in the intraband conductivity formula.

For the $\alpha$-$\mathcal{T}_3$ model, the value of the relaxation time has
not been available yet, but insights can be gained by considering graphene,
where the experimentally measured~\cite{novoselov2004electric} relaxation time
is about $10^{-13}$s. Since the optical response of graphene is appreciable at
high frequencies above $10$THz, a relaxation time on the order of $10^{-13}$s
means that the direction of the field changes much faster than that in the
scattering caused by impurity. In this case, impurity scattering can be
ignored~\cite{jablan2009plasmonics}. If the value of $\omega$ approaches zero,
the conductivity in Eq.~\eqref{eq:2_Intra_final} will diverge, which is
related to the minimal conductivity problem in graphene that remains
unresolved~\cite{sarma2011electronic}. If the value of $\omega$ or $\mu$ is
too large, another difficulty arises for the effectively Hamiltonian. For
graphene in the regime of visible light, it was shown~\cite{stauber2008optical}
that the next nearest neighbor hopping leads to only small corrections to the
Dirac cone, due to the fact that the nearest hopping energy in graphene is
high ($t\approx 2.7$ eV), which corresponds to a photon of frequency of
several hundred Terahertz. For the pseudospin-1 Lieb lattice, experiments
showed that the nearest neighbor hopping energy is smaller than that in
graphene~\cite{slot2017experimental}, but for the dice lattice there has been
no experimental result.

\begin{figure} [ht!]
\centering
\includegraphics[width=\linewidth]{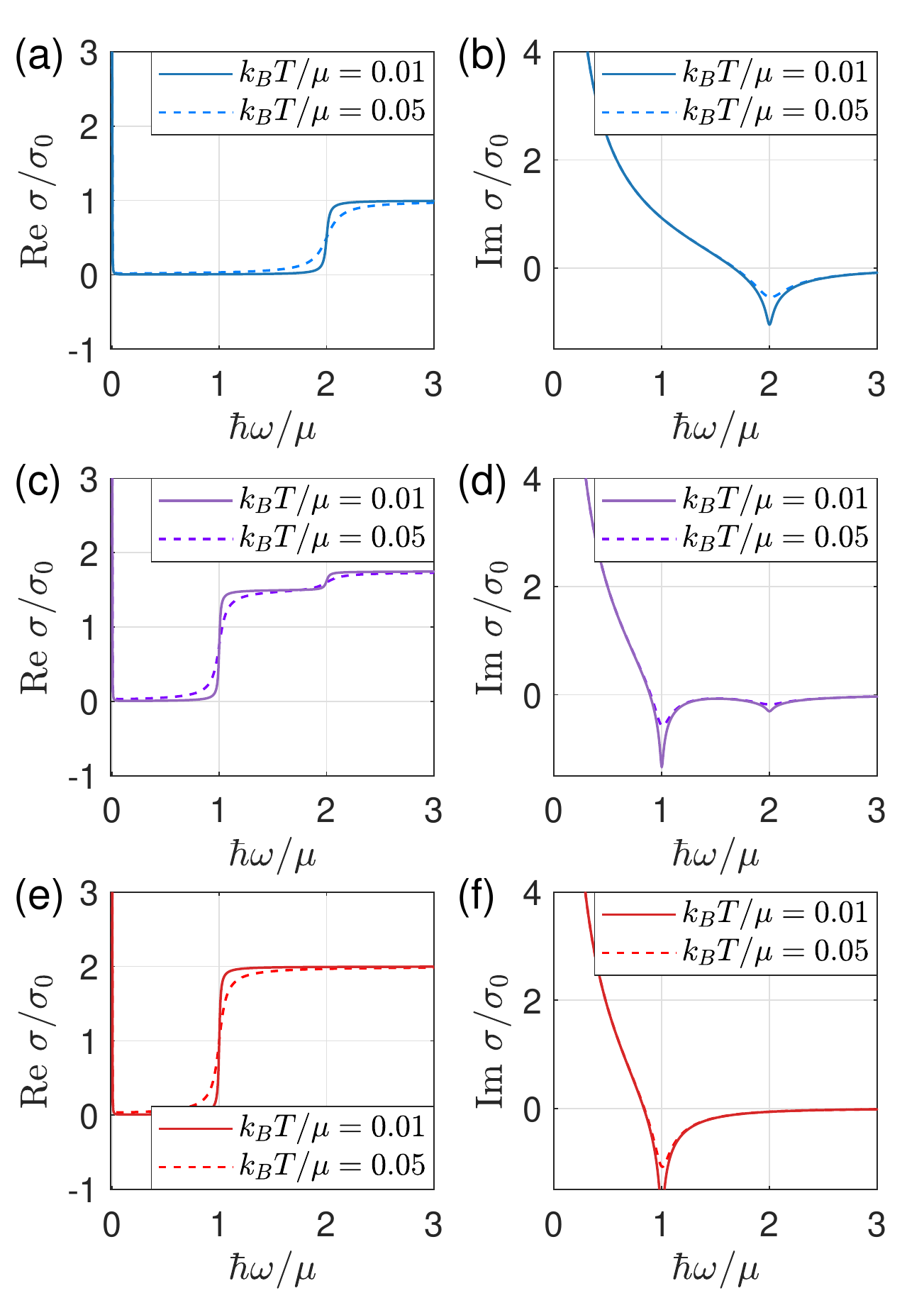}
\caption{Real and imaginary parts of the optical conductivity of the
$\alpha$-$\mathcal{T}_3$ lattice derived from the Kubo formula 
in the absence of any impurity scattering.
(a,c,e) Real part of the optical conductivity for $\alpha=0$ (graphene),
$\alpha=1/\sqrt{3}$, and $\alpha=1$ (pseudospin-1), respectively. In the zero
temperature limit $T\rightarrow 0$, the conductivity is non-zero for
$\hbar\omega/\mu>2$. An interband transition leads to a dip in the
conductivity plot. The step-function type of transition is smoothed out by
finite temperatures. (b,d,f) Imaginary part of the optical conductivity for
$\alpha=0$, $1/\sqrt{3}$, and 1, respectively.}
\label{fig:Kubo}
\end{figure}

Figure~\ref{fig:Kubo} shows the optical conductivity for different values of
$\alpha$ at a finite temperature. The real part of the conductivity is
exemplified in Figs.~\ref{fig:Kubo}(a), \ref{fig:Kubo}(c), and
\ref{fig:Kubo}(e), to which the intraband process has no contribution. There
are two interband transition points, one is the cone-to-cone transition that
occurs for $\hbar\omega/\mu>2$ and another is the flat-band-to-cone
transition that occurs for $\hbar\omega/\mu>1$. (Note that, for $\alpha=0$,
there is no flat band, so the latter transition does not exist.) For
$\alpha=1$, only the flat-band-to-cone transition exists and its magnitude
is twice of that of the cone-to-cone transition for $\alpha=0$. For the
hybrid lattice,  both types of transitions coexist. A finite temperature tends
to smooth the transitions. Figures~\ref{fig:Kubo}(b), \ref{fig:Kubo}(d), and
\ref{fig:Kubo}(f) show the imaginary part of the conductivity, where the
intraband process gives a singularity at $\omega\rightarrow 0$, and each
interband transition leads to a dip for Im $(\sigma) < 0$, which is smoothed
at finite temperatures. A previous study for graphene demonstrated that, when
the imaginary part of the conductivity becomes negative, a new TE mode can
emerge~\cite{mikhailov2007new}. Physically, it would be interesting to
investigate the significance of a negative imaginary part of the optical
conductivity for the $\alpha$-$\mathcal{T}_3$ lattice.

\section{Intrinsic plasmon modes in $\alpha$-$\mathcal{T}_3$ lattice} \label{sec:plane}

An important manifestation of the electromagnetic response of 2D Dirac materials
is the intrinsic plasmon modes. Specifically, for 2D materials on a dielectric
substrate, the Maxwell's equations can be solved analytically, where the
propagating modes follow certain dispersion relationship which depends on the
polarization. A solution method was developed
earlier~\cite{stern1967polarizability,falko1989if} and later adopted
to graphene~\cite{mikhailov2007new,hanson2008dyadic}. The propagating modes
determine the electromagnetic response of the 2D material and they are also
termed ``intrinsic plasmon modes''~\cite{grigorenko2012graphene}. Recently,
the study of the plasmon modes has been extended to other kinds of materials
such as bilayer graphene~\cite{jablan2011transverse}, twisted bilayer
graphene~\cite{stauber2013optical}, silicene~\cite{ukhtary2016broadband}, and
hBN~\cite{musa2017confined}.

Here we study the intrinsic plasmon modes for $\alpha$-$\mathcal{T}_3$ lattice.
Suppose we embed the lattice in-between two materials with dielectric constants
$\epsilon_1$ and $\epsilon_2$, respectively, as illustrated in
Fig.~\ref{fig:schematic}(c). The wavevector $q$ is in the $(x,y)$-plane and
decays along the $z$ direction. Neglecting the thickness of the lattice and
matching the boundary conditions at $z=0$, we obtain the
polarization-dependent (e.g., TM or TE) dispersion
relation~\cite{hanson2008dyadic,ukhtary2016broadband}. In particular,
for the TM polarization, we have
\begin{equation} \label{eq:3_dispersion_TM}
1+\frac{2\pi i \sigma(\omega,\phi)\sqrt{q^2-\omega^2/c^2}}{\omega}=0,
\end{equation}
where $\omega$ is the frequency of the incident field and $c$ is the
speed of light. The solution exists for Im $(\sigma)>0$. For 2D materials, the
conductivity for small incident frequency is described by the Drude
model~\cite{stern1967polarizability,falko1989if}. For TE polarization, the
dispersion relation is
\begin{equation} \label{eq:3_dispersion_TE}
1-\frac{2\pi i \omega\sigma(\omega, \phi)}{c^2\sqrt{q^2-\omega^2/c^2}}=0,
\end{equation}
where the solution exists for Im $(\sigma)<0$. As revealed by previous work
on graphene~\cite{mikhailov2007new}, novel TE polarization can arise for
$1.667<\hbar\omega/\mu<2$. This can be seen from Fig.~\ref{fig:Kubo}(b),
where the interband transition leads to non-zero negative imaginary
conductivity.

TE polarization is a unique feature originated from the linear dispersion
relationship, which arises for relatively high frequencies. Previous work
revealed that the TE transition can be exploited to develop graphene-based
polarizer in the visible light regime~\cite{bao2011broadband} and
high-frequency optical switches~\cite{li2014ultrafast}. The negative imaginary
conductivity leading to TE polarization was exploited to construct a
superscattering system with a curved copper-coated
cylinder~\cite{qian2019experimental}, where a large scattering cross section
was obtained even when the dimension of the scatterer is much smaller than
the wavelength.

The solution of Eqs.~\eqref{eq:3_dispersion_TM} and \eqref{eq:3_dispersion_TE}
exist for nonzero imaginary conductivity, but not all the solutions correspond
to propagating modes. To characterize wave propagation in the lattice, we use
the loss and attenuation length. In particular, the loss is defined as
\begin{equation} \label{eq:3_Loss}
\text{Loss} \equiv \frac{\text{Im} (q)}{\text{Re} (q)},
\end{equation}
which determines the average propagation length in the 2D lattice (a small loss
leads to longer propagation). The attenuation length or skin depth is defined
as~\cite{hanson2008dyadic}
\begin{equation} \label{eq:3_Attenuation}
\xi=\frac{k}{2\pi k_z}=\frac{\lambda_z}{2\pi \lambda}=\frac{\omega/c}{2\pi\text{Re} (\sqrt{q^2-(\omega/c)^2})},
\end{equation}
where $k_z$ is a wave vector measuring the confinement in the $z$ direction
perpendicular to the lattice plane. The strength of the wave in the $z$
direction is proportional to $\exp(-k_z|z|)$ and
\begin{displaymath}
\lambda_z=2\pi/k_z=\text{Re} (2\pi/\sqrt{q^2-(\omega/c)^2}).
\end{displaymath}
A small value of $\lambda_z$ corresponds to strong decay in the $z$ direction
so the wave is well localized in the lattice plane. It is often desired to
have a small loss and strong confinement at the same time, but there is a
trade-off. In general, TM waves tend to have strong confinement with a
large loss, but TE waves have a small loss with weak
confinement~\cite{hanson2008dyadic}.

We calculate the loss and confinement properties for electromagnetic wave
propagation in the $\alpha$-$\mathcal{T}_3$ lattice and compare with those
of graphene. We consider a finite temperature (e.g., $k_BT/\mu=0.01$ or
$k_BT/\mu=0.05$) and make the conductivity dimensionless through
$\tilde{\sigma}=\sigma/\sigma_0$, where $\tilde{\sigma}$ is decomposed into
a real and an imaginary parts:
$\tilde{\sigma}=\tilde{\sigma}'+i\tilde{\sigma}''$. For the TM wave, inserting
the conductivity formula into Eq.~\eqref{eq:3_dispersion_TM}, we get
\begin{equation} \label{eq:3_q_TM}
q_\text{TM}=\frac{\omega}{c}\sqrt{1-\left(\frac{2}{\pi\alpha_0\tilde{\sigma}}\right)^2},
\end{equation}
where $\alpha_0\approx 1/137$ is the fine structure constant. The small
denominator in the second term leads to a large imaginary value of
$q_\text{TM}$, so the loss is large for TM waves. For TE waves, we have
\begin{equation} \label{eq:3_q_TE}
\begin{split}
q_\text{TE}=&\frac{\omega}{c}\sqrt{1-\left(\frac{\pi\alpha_0}{2}\tilde{\sigma}\right)^2}\\
\approx &\frac{\omega}{c}\left[1-i\left(\frac{\pi\alpha_0}{2}\right)^2\tilde{\sigma}'(-\tilde{\sigma}'') \right],
\end{split}
\end{equation}
where $\tilde{\sigma}$ is dimensionless and $\alpha_0$ is small so the
Taylor expansion has been used for the square root. Since the loss is
proportional to $\alpha_0^2$, it is small for TE waves.

\begin{figure} [ht!]
\centering
\includegraphics[width=\linewidth]{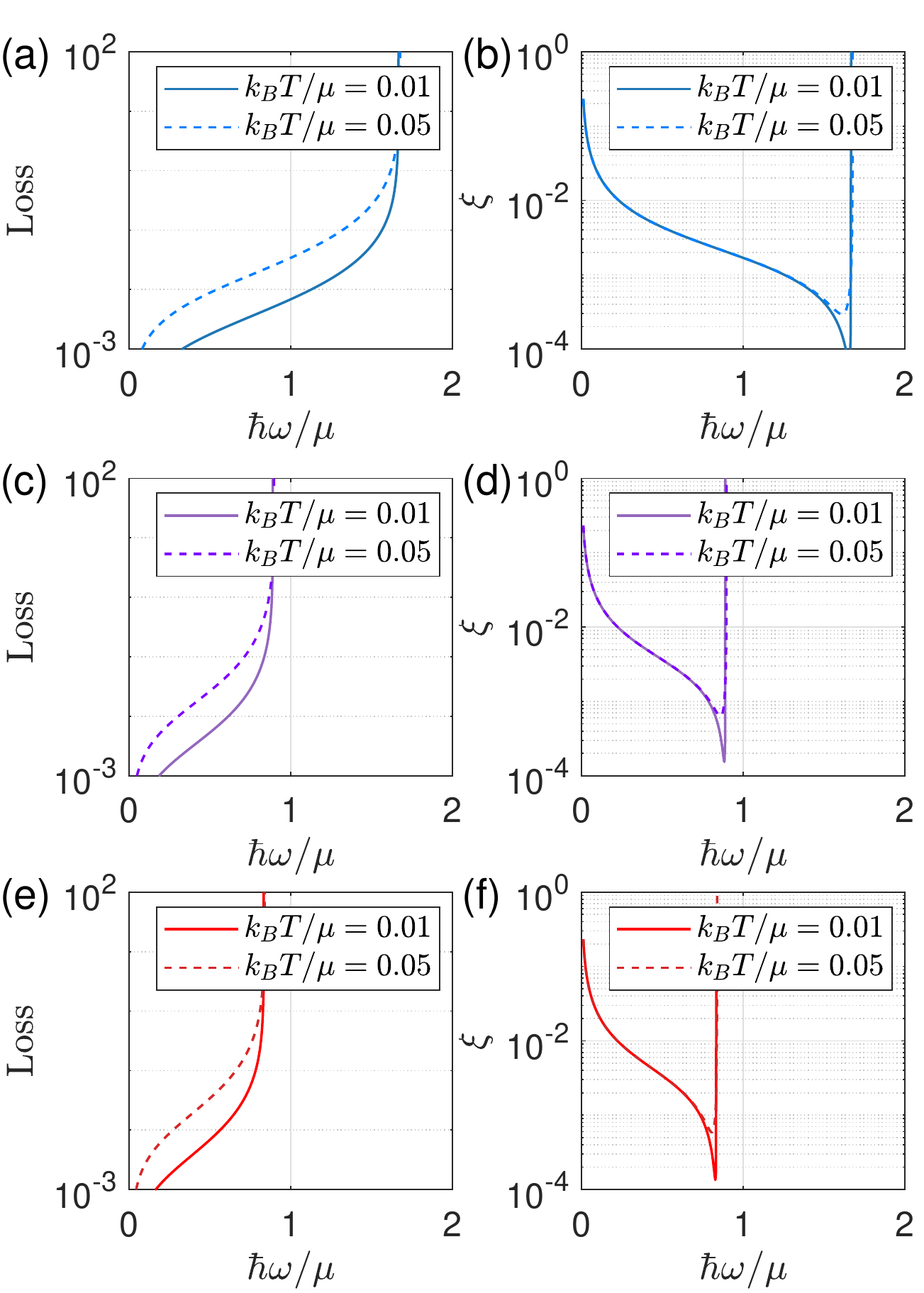}
\caption{Loss and attenuation length for TM wave propagation in the
$\alpha$-$\mathcal{T}_3$ lattice. (a,c,e) Loss versus normalized frequency for
$\alpha=0$, $1/\sqrt{3}$ and 1, respectively. (b,d,f) The respective
attenuation length versus normalized frequency. In each panel, the two curves
correspond to two different temperatures: $T/\mu=0.01$ or $T/\mu=0.05$,
respectively.}
\label{fig:TM}
\end{figure}

The attenuation lengths for the TM and TE waves are
\begin{equation} \label{eq:3_Attenuation2}
\xi_\text{TM}= \frac{\alpha_0}{4} \frac{|\tilde{\sigma}|^2}{\tilde{\sigma}''}
	\ \ \text{and} \ \
\xi_\text{TE}= \frac{1}{\pi^2\alpha_0\tilde{\sigma}''}.
\end{equation}
We first study TM wave propagation in $\alpha$-$\mathcal{T}_3$ lattice for
different $\alpha$ values. Since TM polarization occurs only for
$\mbox{Im}(\sigma)>0$, the underlying waves arise for small $\omega$ values
for which the intraband transition dominates.
Figures~\ref{fig:TM}(a), \ref{fig:TM}(c) and \ref{fig:TM}(e) show the loss
versus the incident frequency for three different $\alpha$ values,
respectively, from Eqs.~\eqref{eq:3_Loss} and \eqref{eq:3_q_TM}. The loss is
small when there is strong intraband transition ($\omega\rightarrow 0$). The
loss becomes large when interband transition is about to happen. As the
temperature increases, the loss become larger but this effect is insignificant.
Figures~\ref{fig:TM}(b), \ref{fig:TM}(d) and \ref{fig:TM}(f) display the
attenuation length for TM waves. From Eq.~\eqref{eq:3_Attenuation2}, we have
that, if the conductivity is purely imaginary, the attenuation length is
proportional to $\tilde{\sigma}''$. As a result, for $\omega\rightarrow 0$
the attenuation length is large. As the temperature increases, the sharp dips
in the attenuation length are smoothed out. As the value of $\alpha$ increases,
the flat-band-to-cone transition begins to dominate, making the imaginary part
of the conductivity negative for small $\omega$, so the viable region for
TM wave propagation becomes smaller.

\begin{figure}
\centering
\includegraphics[width=\linewidth]{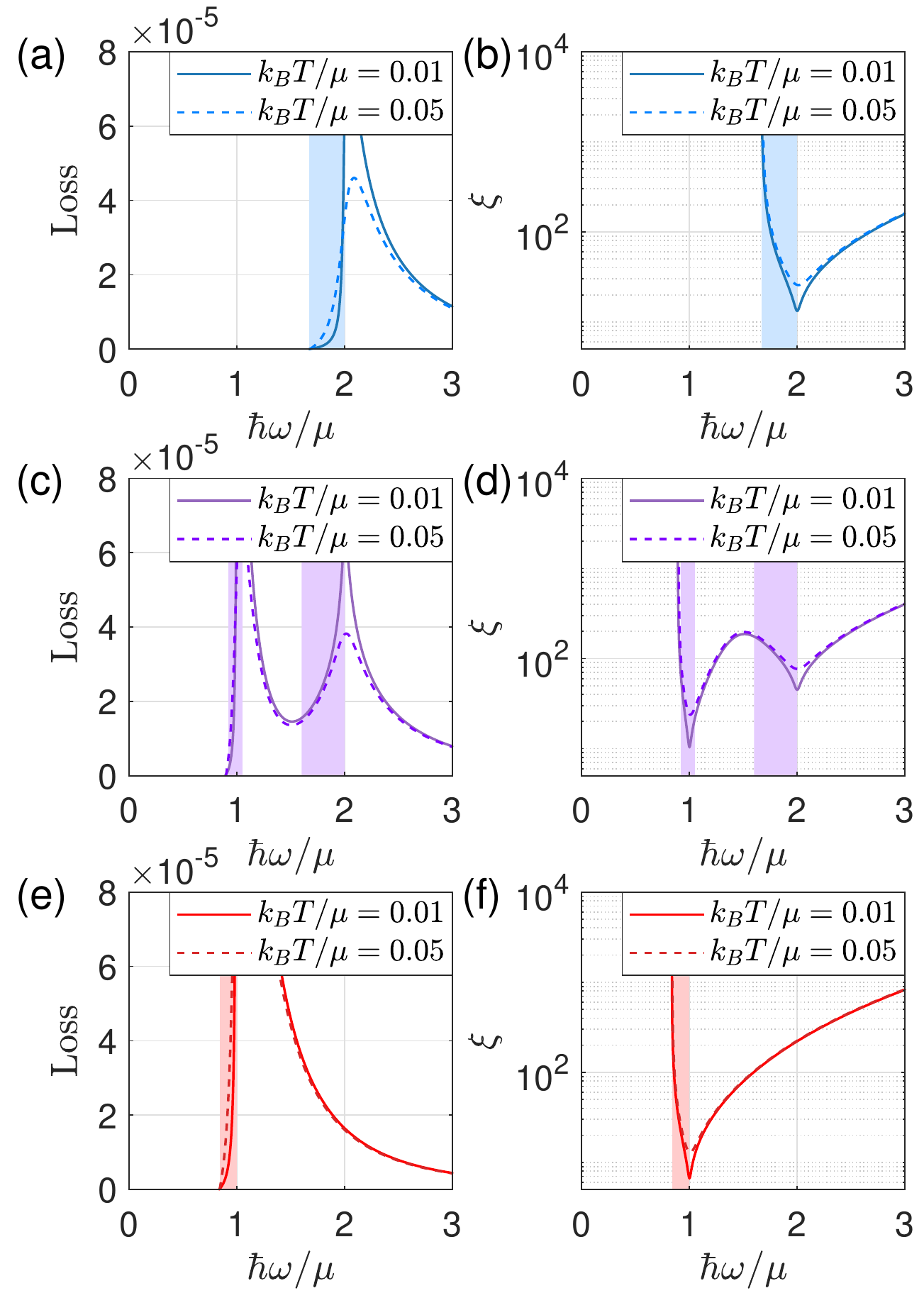}
\caption{Loss and attenuation length for TE wave propagation in
$\alpha$-$\mathcal{T}_3$ lattice. (a,c,e) Loss versus normalized frequency for
$\alpha=0$, $1/\sqrt{3}$ and 1, respectively. (b,d,f) The respective
attenuation length versus the normalized frequency. In each panel, the two
curves correspond to two different temperatures: $T/\mu=0.01$ or $T/\mu=0.05$,
respectively. The shaded regions indicate the possible windows for TE wave
propagation. For $\alpha=0$ (a,b), the region is $\hbar\omega/\mu\in(1.67,2)$.
For $\alpha=1/\sqrt{3}$ (c,d), the region is
$\hbar\omega/\mu\in(0.9,1.1)\cap(1.67,2)$. For $\alpha=1$ (e,f), the region is
$\hbar\omega/\mu\in(0.85,1)$.}
\label{fig:TE}
\end{figure}

Next we study TE wave propagation in the $\alpha$-$\mathcal{T}_3$ lattice
according to Eqs.~\eqref{eq:3_Loss} and \eqref{eq:3_q_TE}.
Figures~\ref{fig:TE}(a), \ref{fig:TE}(c) and \ref{fig:TE}(e) show the loss
versus the incident frequency for three different values of $\alpha$,
respectively, and the corresponding results for the attenuation length are
shown in Figs.~\ref{fig:TE}(b), \ref{fig:TE}(d) and \ref{fig:TE}(f).
The loss is smaller than $10^{-4}$ due to the small imaginary part. After
interband transition arises, the loss increases due to the finite positive
real value of the conductivity. From Eq.~\eqref{eq:3_Attenuation2}, we see
that the attenuation length is inversely proportional to $\tilde{\sigma}''$,
so at the transition point the wave is maximally localized. Note that,
as $\hbar\omega\rightarrow \infty$, even when the loss decreases, the
attenuation length increases, inhibiting the propagating wave. In each panel
of Fig.~\ref{fig:TE}, the shaded regions represent those with relatively
low loss and small attenuation length. For example, for $\alpha=0$, this occurs
before the cone-to-cone transition, i.e., for $\hbar\omega/\mu\in(1.67,2)$,
which agrees with the previous result for graphene~\cite{mikhailov2007new}.
For $\alpha=1/\sqrt{3}$, there are two shaded regions due to the coexistence
of two transitions. For $\alpha=1$, there is one shaded region due to the
flat-band-to-cone transition. A higher temperature can lead to an increase in
both the loss and attenuation length.

To summarize these results briefly, we have that, for TM waves, the loss is
high - typically about $0.1$ to $1$ but the attenuation length is small, so
the TM waves are highly localized with high loss. Because of the positive
imaginary part of the conductivity $\sigma(\omega,\phi)$, as $\alpha$
increases the propagation region shrinks. For TE waves, the loss is low -
typically less than $10^{-4}$ but the attenuation length is large, so the
waves are weakly localized. As $\alpha$ increases from zero, the bandwidth
for propagation first increases due to the coexistence of multi-interband
processes. For $\alpha=1$, the flat-band-to-cone transition has a large
magnitude and is the only process present.

There are two unique features of electromagnetic wave propagation in the
$\alpha$-$\mathcal{T}_3$ lattice. First, comparing panels (a,b) with (e,f) in
Fig.~\ref{fig:TE}, we have that the TE waves for pseudospin-1 lattice have
a narrow frequency width and a small attenuation length, indicating that
the intrinsic plasmon mode is strongly localized, which is due to the
transition from the flat-band to the linear band. Second, Figs.~\ref{fig:TE}(c)
and \ref{fig:TE}(d) indicate that multiple frequency TE polarization waves
can arise: one is the same as graphene at $\hbar\omega/\mu\approx2$ due to
the cone-to-cone transition and another occurs at $\hbar\omega/\mu\approx1$,
which is due to the flat-band-to-cone transition. For $\mu=0.5 $ eV, the
linear dispersion relation holds, so the regions that can support TE wave
propagation correspond to 110--125 THz and 190--240 THz at the room temperature
($k_BT/\mu=0.05$ - about $300$K).

\begin{figure} [ht!]
\centering
\includegraphics[width=\linewidth]{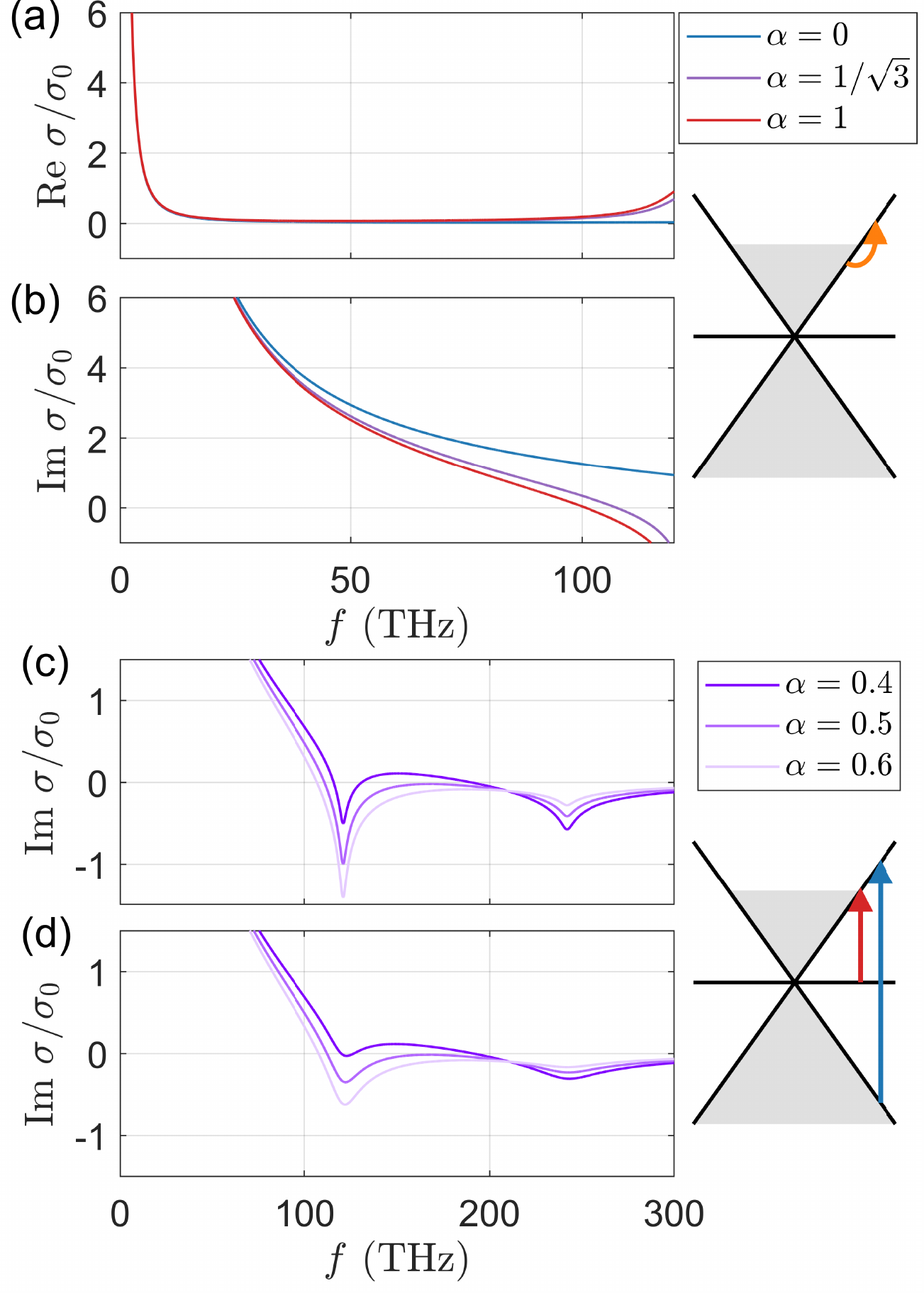}
\caption{
Emergence of TE polarization waves in $\alpha$-$\mathcal{T}_3$ materials.
(a,b) Real and imaginary parts of the optical conductivity, respectively,
for $\alpha = 0$ (graphene), $\alpha = 1/\sqrt{3}$, and $\alpha = 1$
(pseudospin-1). The parameters are $\mu=0.5$eV, $T=300$K and
$\tau=6.4\times 10^{-13}$. In the low frequency regime $\omega\rightarrow 0$,
intraband transitions dominate so the conductivity for different values of
$\alpha$ converges. (c,d) Imaginary part of the optical conductivity for the
hybrid case ($\alpha\ne 0$) at two temperatures ($T=60$K and $T=300$K,
respectively) for $\mu=0.5$eV. The two local minima are associated with two
different interband transitions, which broaden as the temperature increases.}
\label{fig:5}
\end{figure}

We have studied the effect of impurity scattering on the optical conductivity.
Figures~\ref{fig:5}(a) and \ref{fig:5}(b) display the conductivity curve in
the frequency range $[0,300]$ THz in the presence of finite impurity
scattering. The value of the relaxation time is taken to be that of graphene
($\alpha=0$)~\cite{jablan2009plasmonics}, as the experimental value of this
time for $\alpha\ne 0$ has not been available yet. It can be seen that finite
impurity scattering can change the real part of the intraband conductivity and
smooth out the $\delta$-function. However, the change mainly occurs for
frequencies below 10THz and is generally insignificant. In fact, for
frequencies above 100THz, finite impurity scattering has little effect on the
optical conductivity. Our formula of the intraband conductivity also suggests
that it depends on the relaxation time but not on the value of $\alpha$. We
note that a similar observation was made earlier~\cite{illes2015hall}.

The interval in $\alpha$ in which the two TE polarization waves arise can be
seen from Figs.~\ref{fig:5}(c) and \ref{fig:5}(d). At low temperatures, there
are two distinct local minima generated by interband transitions for
$\alpha\in[0.4,0.6]$. When the temperature increases to $k_BT/\mu=0.05$,
the interval shrinks. This phenomenon can be understood, as follows. According
to Eqs.~\eqref{eq:2_Inter1_final} and \eqref{eq:2_Inter2_final}, the strength
of the cone-to-cone transition is proportional to $\cos^2(2\phi)$ and that of
the flat-band-to-cone transition is proportional to $2\sin^2(2\phi)$, where
$\alpha\equiv\tan(\phi)$. Without taking into consideration intraband
transitions, the two types of interband transition would have the same
magnitude for $2\sin^2(2\phi)=\cos^2(2\phi)$, which correspond to
$\alpha \approx 0.3$. From Eq.~\eqref{eq:2_Intra_final}, intraband transitions
give a positive imaginary value in the conductivity and the interband
transitions generate a negative imaginary value. Since the strength of
intraband transitions is inversely proportional to $\omega$, the effective
strength of the flat-band-to-cone transition is reduced. To have approximately
the same strength for the cone-to-cone and flat-band-to-cone transitions, it is
necessary to increase the value of $\alpha$ to facilitate the flat-band-to-cone
transition. Numerically, the optimal $\alpha$ interval in which the two types
of interband transition are approximately equal can be obtained by
monitoring the respective local minima in the conductivity generated by the
transitions and examining when the two local minima reach a similar magnitude.
From Response Figs.~\ref{fig:5}(c) and \ref{fig:5}(d), we note that the two
local minima are similar for $\alpha\in(0.4,0.6)$. At finite temperatures,
the local minima will be smoothed out, so the interval shrinks as the
temperature increases.

\section{Resonant scattering from an $\alpha$-$\mathcal{T}_3$ lattice coated dielectric sphere} \label{sec:sphere}

In electromagnetics, for applications such as optical sensing, imaging, tagging
and spectroscopy~\cite{danilov2000detection,del2011octave,schliesser2012mid},
enhanced scattering is desired. There are also applications where reduced
scattering is sought, such as cloaking~\cite{miller2006perfect,alu2009cloaking},
which can be realized through the technique of scattering cancellation. In
particular, by coating an additional material layer outside the original
scatterer~\cite{alu2005achieving}, destructive interference can be induced
between the scattering waves from the two structures. Not only is the method
of coating capable of inducing cloaking, but it can also enhance
scattering~\cite{peurifoy2018nanophotonic} with proper coating materials and
design. For example, a dielectric structure coated with graphene can lead to
cloaking at some wavelength~\cite{chen2011atomically,zhao2015resonance,
farhat20133d,monticone2013cloaked}, but at other wavelength optical scattering
can be enhanced~\cite{li2015tunable}. These behaviors can be controlled by
tuning the chemical potential.

For scattering from a graphene coated structure, the surface conductivity
will generate certain polarization. The change in the scattering cross section
is strongly related to the intrinsic plasmon modes. For plasmon modes with TM
polarization, superscattering or cloaking typically arise in the 1--10 THz
range due to intraband transitions~\cite{chen2011atomically,zhao2015resonance,
farhat20133d}. For TE waves, the frequency is higher and is strongly modulated
by the temperature~\cite{li2015tunable}. Due to the small imaginary part of the
conductivity, scattering is typically weak. Multilayer structures can be used
to enhance the polarization~\cite{bao2011broadband,musa2017confined,
qian2018multifrequency}. A previous experiment demonstrated that, for graphene,
scattering structures with up to five layers can be designed, with scattering
loss comparable to that with the monolayer structure~\cite{bao2011broadband}.

Here we study optical scattering from a dielectric sphere coated by multiple
layers of $\alpha$-$\mathcal{T}_3$ lattice. For a multilayer lattice, the
dielectric constant (relative permittivity) is defined
as~\cite{bao2011broadband,farhat20133d}
\begin{equation}
\epsilon_{\alpha,N}=1+i\frac{N\sigma(\omega,\phi)}{r_2\epsilon_0\omega},
\end{equation}
where $N$ is the number of layers, $\epsilon_0$ is vacuum permittivity, and
$r_2$ is the total depth of the coating structure whose value is usually
chosen to be much smaller than the device dimension to reduce the finite-depth
effect~\cite{farhat20133d}. We choose a dielectric sphere of radius $r_1=100$
nm with dielectric constant $\epsilon_1=2.1$, corresponding to materials such
as Polytetrafluoroethylene (PTFE)~\cite{ehrlich1953dielectric}, and set
$r_2=1 \ \text{nm} \ll r_1$. Computationally, the multilayer lattice structure
can be treated as a single layer, as all that is required for calculating the
electromagnetic scattering cross section is the conductivity of the
$\alpha$-$\mathcal{T}_3$ lattice. Once the dimension and frequency dependent
dielectric constant $\epsilon_{\alpha,N}$ is given, the scattering system can
be analytically solved with proper solutions of the Maxwell's equations.

\begin{figure} [ht!]
\centering
\includegraphics[width=\linewidth]{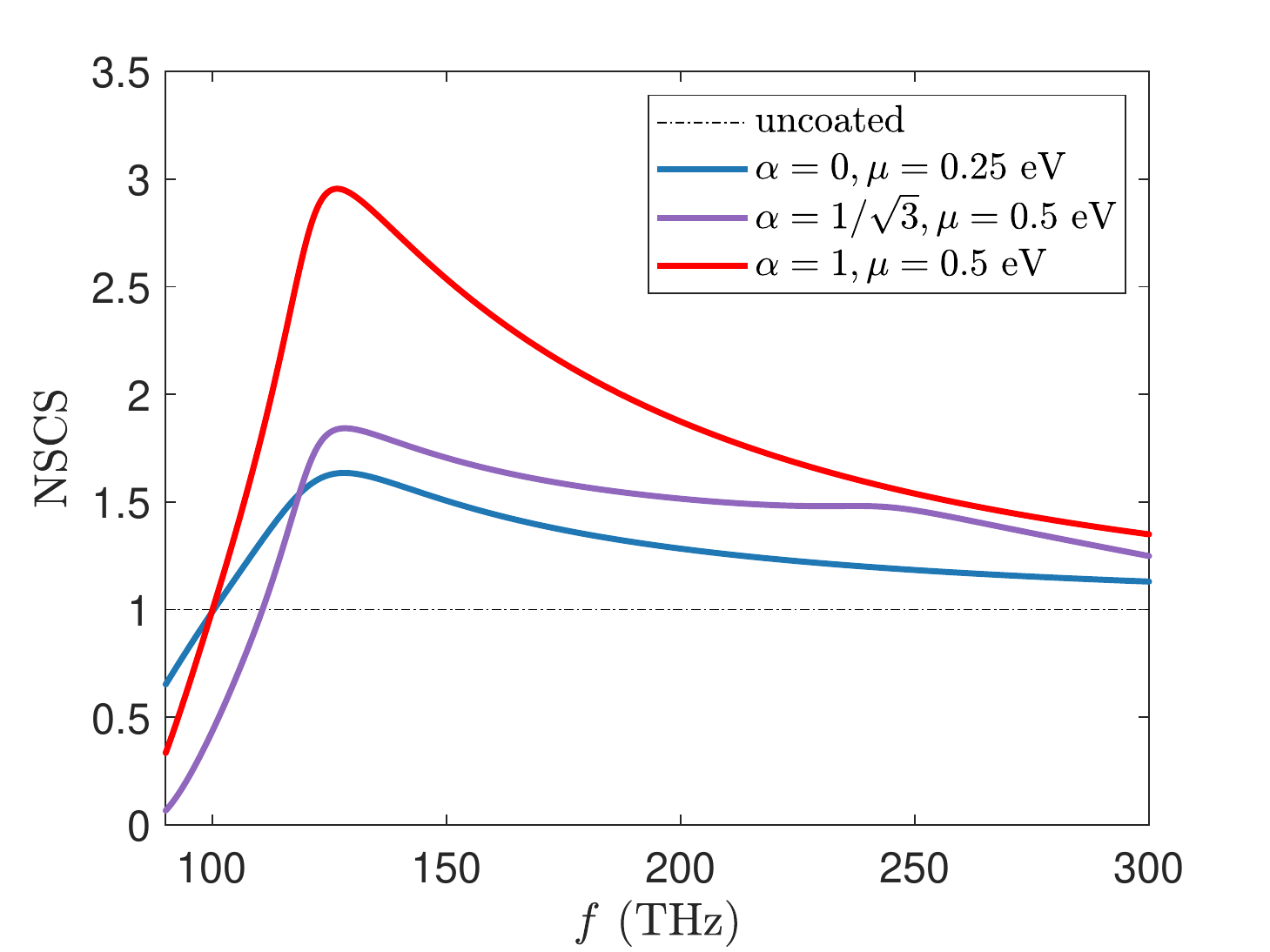}
\caption{ Scattering cross section from a dielectric sphere coated with
multiple layers of $\alpha$-$\mathcal{T}_3$ lattice. Shown is the ratio
between the scattering cross sections with and without coating ($\sigma$ and
$\sigma_0$, respectively) versus frequency. For $\alpha=0$ there is only one
resonance peak at $f\approx130 $ THz. For $\alpha=1/\sqrt{3}$, there are two
peaks at $f\approx130$ and $f\approx 260$, respectively. For $\alpha=1$ there is
only one peak at $f\approx130$ and the value of NSCS nearly doubles as
compared with the case of graphene coating.}
\label{fig:NSCS}
\end{figure}

\begin{figure} [ht!]
\centering
\includegraphics[width=\linewidth]{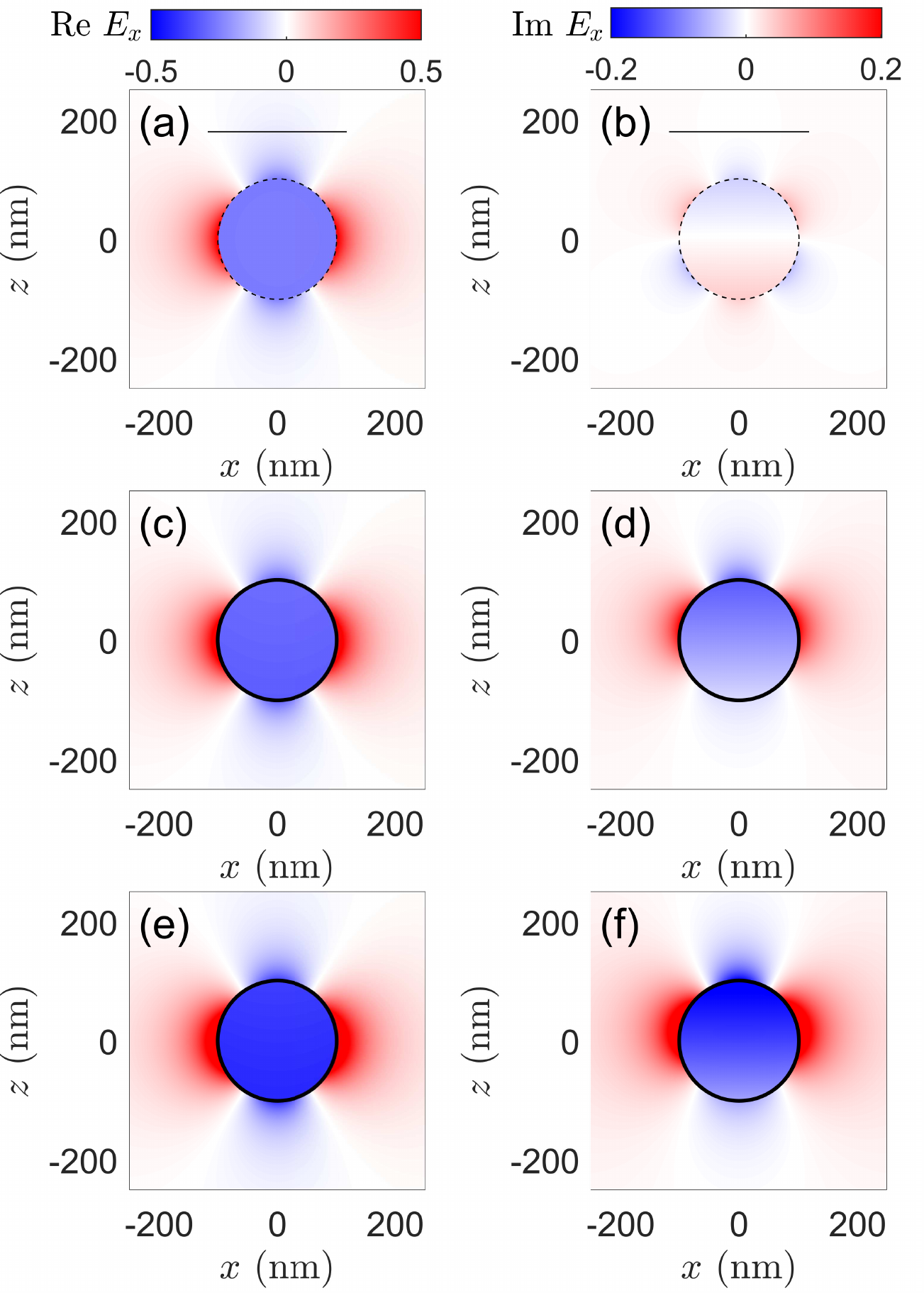}
\caption{Examples of scattering wavefunction. Shown are the real (left column)
and imaginary (right column) parts of the scattering field for three cases:
(a,b) without coating where the horizontal line segment has the length
$\lambda/10$ and indicates that the wavelength is much larger than the size of
the scatterer, (c,d) graphene coating and (e,f) pseudospin-1 lattice coating.
The scattering field is relatively strong with the pseudospin-1 lattice
coating.} 	
\label{fig:SF}
\end{figure}

The incident wave is chosen to propagate in the $z$ direction and polarized
along the $x$ direction, and the scattering wave is calculated by using
the transfer matrix method~\cite{hulst1981light,johnson1996light}. After
decomposing the scattering wave into a series of spherical Bessel functions,
the coefficient for each basis is obtained. The scattering cross section
is given by
\begin{equation}
\sigma_\text{sc}=\sum_i (2i+1) (|c^\text{TM}_i|^2+|c^\text{TE}_i|^2)
\end{equation}
where $c^\text{TM}_i$ and $c^\text{TE}_i$ are the coefficients for scattering
wave with the corresponding polarization. (The calculation details are
presented in Appendix~\ref{Appendix_C}.) To characterize the change in the
scattering cross section before and after coating, we define the normalized
scattering cross section as the ratio between the cross sections with and
without the coated structure:
\begin{equation} \label{eq:4_NSCS}
\text{NSCS}=\frac{\sigma_\text{sc coated}}{\sigma_\text{sc uncoated}}.
\end{equation}
Figure~\ref{fig:NSCS} shows NSCS versus frequency for different scattering
structures. For a meaningful comparison, we choose the number of layers to
be $N=5$ and set the temperature to be $T=300$ K for all structures. Without
coating, the value of NSCS is one. For $\alpha=0$ (graphene), when the chemical
potential of $0.25$ eV is applied, there is a resonance at $f\approx130$ THz,
which is approximately $\hbar\omega\approx2\mu$ and corresponds to a TE
intrinsic plasmon mode in graphene. For $\alpha=1/\sqrt{3}$, two resonance peaks
arise but their strength is weak, which correspond to the two different band
transition processes. Similar to the multiple frequency phenomenon associated
with propagating wave, the resonances are temperature-dependent. For
$\alpha=1$, we double the chemical potential for comparing with the case of
graphene, which results in a resonance at the same frequency. The resonance
for $\alpha=1$ can be attributed to the flat-band-to-cone transition, which
is strong due to the large imaginary part of the conductivity.

Figure~\ref{fig:SF} shows the scattering field for $f=127$ THz, corresponding
to the left resonant peak in Fig.~\ref{fig:NSCS}, for three cases: without
coating, graphene coating, and pseudospin-1 lattice coating. Without coating,
as shown in Figs.~\ref{fig:SF}(a) and \ref{fig:SF}(b), the scattering field is
weak because the wavelength is much larger than the dimension of the scatterer.
Figures~\ref{fig:SF}(c) and \ref{fig:SF}(d) are for $\alpha = 0$ with the
same parameter setting as in Fig.~\ref{fig:NSCS}. The scattering field for
$\alpha = 1$ is shown in Figs.~\ref{fig:SF}(e) and \ref{fig:SF}(f),
demonstrating that the field is enhanced as compared with the case of graphene
and greatly enhanced as compared with the case without coating.

\section{Discussion} \label{sec:discussion}

Two-dimensional Dirac materials with a flat band have become an active frontier
in condensed matter physics and materials science, and it is of interest to
understand the optical responses of these materials. In this regard, the basic
physical quantity is the optical conductivity. Previous studies focused on the
real part of the conductivity. Utilizing $\alpha$-$\mathcal{T}_3$ lattice as a
vehicle, we have derived a complete formula for the optical conductivity which
includes both the real and imaginary parts. The base of our derivation is the
Kubo formula and we have also exploited the Kramers–Kronig relation to provide
an alternative derivation that leads to the same conductivity formula.
Physically, the presence of a flat band, together with the Dirac cone pairs,
gives rise to richer transitions as compared with Dirac materials without a
flat band, such as graphene. We have demonstrated that the conductivity of
$\alpha$-$\mathcal{T}_3$ lattice has three components, each contributed to by
a distinct optical transition. Specifically, the contribution from intraband
transitions is similar to that for Dirac materials without a flat band. While
the cone-to-cone transition is pronounced for $\alpha = 0$ (graphene), it
weakens as $\alpha$ increases towards one. For $\alpha > 0$, the
flat-band-to-cone transition is dominant. For $\alpha = 1$ where the influences
of the flat band are the strongest possible, the contribution to the
conductivity from the cone-to-cone transition becomes negligible as compared
with that from the flat-band-to-cone transition.

With the derived frequency-dependent conductivity formula for
$\alpha$-$\mathcal{T}_3$ lattice, we have investigated two problems
that are fundamental for developing $\alpha$-$\mathcal{T}_3$ based optical
devices. The first is electromagnetic wave propagation, which results in
intrinsic plasmon modes in the $\alpha$-$\mathcal{T}_3$ lattice, whose physical
properties depend on the polarization. Our calculations have revealed that TM
polarized waves are the result of intraband transitions, which occur in the
frequency range 1--10 THz. These waves also arise in other 2D Dirac materials
such as graphene. In contrast, TE polarized waves are generated by interband
transitions, which arise in a higher frequency range: 100--300 THz. For
$0 < \alpha < 1$, two interband transitions occur simultaneously, which
generate two TE surface waves at $\hbar\omega/\mu\approx 1, 2$, respectively.
This is verified by calculating the loss and attenuation length that are
minimized in the frequency region. The second problem is scattering from a
dielectric sphere coated with multiple layers of $\alpha$-$\mathcal{T}_3$
material. The finding is that, due to a reduction in the imaginary part of
the optical conductivity at finite temperatures, the scattering of TE polarized
waves is weaker than that of TM waves. The multilayer scattering structure can
then be used to enhance certain
polarization~\cite{bao2011broadband,musa2017confined,qian2018multifrequency}.

For 2D Dirac materials, an approach to generating scattering or transport
behaviors at multiple frequencies is to use some structure with a special
band, such as bilayer graphene with four bands~\cite{jablan2011transverse}.
For a proper choice of the chemical potential, the phenomenon of frequency
doubling can occur. In materials similar to graphene such as
silicene~\cite{ukhtary2016broadband}, the sublattice and valley symmetries
are broken, thereby generating multiple energy bands. For these materials,
the electromagnetic property depends on the chemical potential, so the
intrinsic plasmon modes depend on the chemical potential as well. The
frequency tunability of these materials is typically weaker than that of
$\alpha$-$\mathcal{T}_3$ materials for $\alpha \alt 1$. Recently there is a
growing interest in hexagonal boron nitride (hBN)
materials~\cite{qian2018multifrequency} whose hyperbolic permittivity
can lead to a multi-frequency resonant scattering. However, for the TE
polarization in this material, the resonant frequency is fixed. In contrast,
our study here has demonstrated that, for the $\alpha$-$\mathcal{T}_3$
lattice, insofar as the linear dispersion holds, TE polarized waves can be
tuned by adjusting the chemical potential,

A general result is that the optical responses of $\alpha$-$\mathcal{T}_3$
materials for $\alpha \alt 1$ are in general more pronounced than those for
$\alpha \agt 0$, e.g., graphene, as the conductivity due to the
flat-band-to-cone transition is twice of that due to the cone-to-cone
transition. For the same frequency, this means that for pseudospin-1
materials, the chemical potential is effectively doubled, making the optical
responses twice as strong as those for graphene. A physical reason for this
enhancement is that the plane wave in the pseudospin-1 lattice has a smaller
attenuation length due to the large imaginary part of the optical conductivity
as compared to that in graphene. From the point of view of resonant scattering,
at the same frequency, a larger scattering cross section can arise in
pseudospin-1 materials in comparison with graphene. We note that hBN materials
can also generate stronger surface plasmon waves than
graphene~\cite{woessner2015highly, musa2017confined}.

When performing the scattering calculation for a multilayer
$\alpha$-$\mathcal{T}_3$ structure, we ignored the effect of interlayer
coupling. This approximation can be justified, as follows. In a study of
the optical conductivity in monolayer and double-layer $\alpha$-$\mathcal{T}_3$
lattice~\cite{iurov2020many}, it was observed that the coupling decays
exponentially with the interlayer spacing. The results on the plasmon density
indicate that only for small layer spacing will the original peak split into
two distinct peaks. For reasonably large spacing, e.g., $d=5 k_F^{-1}$, the
layer coupling effect can be neglected.

An extensively studied case is graphene, where monolayer graphene has two
linear bands. The bilayer graphene system studied in
Refs.~\cite{abergel2007optical,jablan2011transverse}
has four bands due to interlayer coupling and, as a result, there is
splitting in the plasmon peaks. However, when the interlayer distance is
large, the coupling effect becomes negligible. For $v_F=10^6$ and $\mu=1$ eV,
this critical spacing is about $d=3.3$ nm. In experiments,
stacking geometry is often used to generate multilayer graphene system,
where the interlayer coupling effect is insignificant and the electronic
structure of the system is similar to that of monolayer
graphene~\cite{bao2011broadband}.

We remark on the possible many-body effects. In graphene, the long-range
Coulomb interaction can lead to a renormalized Fermi velocity for $\mu=0$ and
$T=0$. In Ref.~\cite{stauber2017interacting}, a Hartree-Fock theory was
developed with no fitting parameters but with a topological invariant. A
theoretical calculation of the Fermi velocity renormalization agreed with the
experimental data. While the renormalization can change the optical
conductivity, the effect is small.

In Ref.~\cite{hausler2015flat}, a similar renormalization effect was observed
in lattices with a flat band. For $\mu=0$ and a partially filled flat band,
the Drude weight has several zeros depending on the inverse ratio of the band
filling. Since the intraband conductivity is proportional to the Drude weight,
such a change may also lead to conductivity oscillations for $\mu=0$ and small
$\omega$. Our work focuses on the case of a positive chemical potential $\mu$,
where the renormalization effect is small.

Can the flat band contribute to the intraband conductivity? In a previous
work~\cite{gneiting2018lifetime}, the authors considered a lattice
model with a flat band in the presence of correlated disorders that provide
coupling between the flat-band and dispersion band states. For a finite
lattice and an initial Gaussian wave packet in the real space, some states in
the flat band are unoccupied, giving rise to intraband transitions within
the flat band. In our study, correlated disorders were assumed to be absent
and the Fermi energy is positive so that the flat band is fully occupied
for $T\rightarrow 0$. Since only the states near the Fermi surface are excited, the flat band gives no contribution to the intraband conductivity. The same
observation was made in another previous study~\cite{tabert2016optical},
where the real part of the conductivity was derived with the finding that the
intraband conductivity does not depend on $\alpha$.

With regard to impurity scattering, previously the imaginary part of 
conductivity was studied in the low frequency 
regime~\cite{marsiglio1996imaginary,jiang1996imaginary,pronin2001optical}. 
It was found that the relaxation time $\tau$ has a significant effect on the 
convergence of the conductivity. For a small impurity scattering rate, the 
imaginary part tends to diverge for near zero frequency. The values of the 
relaxation time for flat-band Dirac materials are not known at the present. 
However, our study focuses on the physically more relevant high frequency 
regime.

To summarize, the analysis of our complete formula of the optical conductivity
for $\alpha$-$\mathcal{T}_3$ suggest a number of phenomena that can be useful
for designing optical devices based on this type of generalized 2D Dirac
materials. For example, for $\alpha\in(0.4, 0.6)$, because of the coexistence
of cone-to-cone and flat-band-to-cone transitions, the corresponding lattice
can be exploited for TE wave based broad-band devices. Another phenomenon is
that the magnitude of the optical conductivity of pseudospin-1 materials is
twice as large as that of graphene due to a reduction in the energy required
for a flat-band-to-cone transition. A strong TE wave can then be generated and
sustain at $\hbar\omega/\mu\approx1$, giving the possibility to design optical
devices using $\alpha$-$\mathcal{T}_3$ ribbon or other coupling
structures~\cite{nikitin2011edge, zhao2015resonance,meng2020ultrathin}.
Recent work in quantum plasmonics has suggested that edge states in graphene
can lead to a blue shift in the plasmon modes~\cite{wedel2018emergent}.
It would be interesting to exploit $\alpha$-$\mathcal{T}_3$ materials for
applications in quantum plasmonics.

\section*{Acknowledgement}

This work was supported by AFOSR under Grant No.~FA9550-21-1-0186.

\appendix

\section{Optical matrix for $\alpha$-$\mathcal{T}_3$ lattice} \label{Appendix_A}

In the tight-binding framework, the low energy excitations in the
$\alpha$-$\mathcal{T}_3$ lattice are described by the Hamiltonian
\eqref{eq:2_Hamiltonian}. Associated with the first valley ($v=1$), there are
three bands: $0,\pm 1$, corresponding to the flat band, conduction and valence
bands, respectively. The eigenfunctions are
\begin{equation}
|\psi_{\pm 1}\rangle=\frac{1}{\sqrt{2}} \begin{pmatrix} \cos \phi e^{i\theta_\mathbf{k}} \\ \pm 1 \\
\sin\phi e^{-\theta_\mathbf{k}}\end{pmatrix}
\end{equation}
and
\begin{equation}
|\psi_{0}\rangle=\begin{pmatrix} \sin\phi e^{i\theta_\mathbf{k}} \\ 0 \\
-\cos\phi e^{-i\theta_\mathbf{k}} \end{pmatrix},
\end{equation}
where $f_\mathbf{k}=|f_\mathbf{k}|e^{i\theta_\mathbf{k}}$ and
$\theta_\mathbf{k}$ is the angle of $\mathbf{k}$ in the polar coordinates.
For the second valley, we have $f_{\mathbf{k},v=-1}=-f_{\mathbf{k},v=1}^*$,
so the solutions can be obtained from a sign change:
$\theta_\mathbf{k}=-\theta_\mathbf{k}$. Because of the mirror symmetry in
the integration with respect to $\mathbf{k}$, the minus sign will not change
the results.

The optical matrix elements associated with the current operator in the $x$
direction are given by
\begin{equation} \label{eq:A_matrix}
\begin{split}
|\langle \mathbf{k}, \pm | j_x| \mathbf{k}, \pm \rangle |^2 &=e^2v_F^2\cos^2\theta_\mathbf{k} \\
|\langle \mathbf{k}, \pm |j_x| \mathbf{k}, \mp \rangle |^2&=e^2v_F^2\sin^2\theta_\mathbf{k}\cos^2(2\phi)\\
|\langle \mathbf{k}, 0 |j_x|\mathbf{k}, \pm\rangle |^2 &=|\langle \mathbf{k}, \pm |j_x|\mathbf{k}, 0\rangle |^2\\
&=\frac{e^2v_F^2}{2}\sin^2\theta_\mathbf{k}\sin^2(2\phi).
\end{split}
\end{equation}

\section{Optical conductivity of $\alpha$-$\mathcal{T}_3$ lattice} \label{Appendix_B}

\subsection{Derivation based on the Kubo formula} \label{Appendix_B1}

With the Kubo formula~\eqref{eq:2_Kubo}, the summation over different states
can be simplified to the summation from $k=k'$, which corresponds to the direct
optical transition. There are three types of band transitions.

\paragraph*{Intraband transition.}
The transition is from the conduction band to itself with
$E_n-E_m\rightarrow 0$ and $E_n\approx E_m \approx \mu$. Under the
approximations, the Fermi-Dirac distribution can be written as
\begin{equation}
\frac{F(E_m)-F(E_n)}{E_n-E_m}=-\left. \frac{\partial F}{\partial \epsilon}\right|_{\epsilon=\mu}=\delta(\epsilon-\mu).
\end{equation}
Equation~\eqref{eq:2_Kubo} then becomes
\begin{equation} \label{eq:B_Intra1}
\sigma^{(1)}(\omega,\phi) =\frac{\hbar}{i\pi^2}\iint dk_xdk_y\frac{\partial F}{ \partial \epsilon} \frac{j_{nm}^2}{\hbar\omega}.
\end{equation}
Inserting the optical matrix elements into Eq.~\eqref{eq:A_matrix} and using
the linear dispersion relationship $E=\hbar v_F|\mathbf{k}|$, in the polar
coordinates we have
\begin{equation} \label{eq:B_coordinate}
\iint dk_xdk_y j_{nm}^2=\frac{e^2}{\hbar^2}\int_0^\infty \epsilon d\epsilon \int_0^{2\pi} \cos^2\theta_\mathbf{k} d\theta_\mathbf{k}.
\end{equation}
Equation~\eqref{eq:B_Intra1} becomes
\begin{equation}
\sigma^{(1)}(\omega,\phi)=\frac{e^2}{i\pi\hbar^2\omega}\int \epsilon [-\delta(\epsilon-\mu)] d\epsilon=\frac{ie^2\mu}{\pi\hbar^2\omega}.
\end{equation}
Introducing $\sigma_0=e^2/(4\hbar)$, we obtain the intraband conductivity as
\begin{equation} \label{eq:B_Intra_final}
\sigma^{(1)}(\omega,\phi)=\frac{4i\mu\sigma_0}{\pi\hbar\omega}.
\end{equation}

\paragraph*{Cone-to-cone transition.}
This transition occurs from $|-\rangle$ to $|+\rangle$ and vice visa. Due to
the involvement of two different bands, an additional factor of $2$ arises
in the summation;
\begin{equation}
\begin{split}
\sigma^{(2)}(\omega,\phi) =&\frac{\hbar}{i\pi^2}\sum_{n,m}\frac{F(E_m)-F(E_n)}{E_n-E_m}\times\\
&\frac{j_{nm}^2(-2\hbar\omega)}{(\hbar\omega)-(E_n-E_m)^2}.
\end{split}
\end{equation}
Since $\mathbf{k}=\mathbf{k}'$ and because $E_n$ and $E_m$ belong to different
bands, we can write $E_n=\epsilon$ and $E_m=-\epsilon$. Using the integral
in Eq.~\eqref{eq:B_coordinate} and the optical matrix elements
Eq.~\eqref{eq:A_matrix}, we get
\begin{equation} \label{eq:B_Inter1_int}
\begin{split}
\sigma^{(2)}(\omega,\phi)=&\cos^2(2\phi)\frac{e^2}{i\pi\hbar}\int [F(-\epsilon)-F(\epsilon)]\times\\
&\frac{\hbar\omega}{4\epsilon^2-(\hbar\omega)} d\epsilon.
\end{split}
\end{equation}
The difference in the Fermi-Dirac distribution from the case of intraband
transition implies that a non-zero value occurs only for $\epsilon>\mu$ or
$\epsilon<-\mu$, so in the polar coordinates only the first term is meaningful.
We obtain
\begin{equation}
\sigma^{(2)}(\omega,\phi)=\cos^2(2\phi)\frac{e^2}{i\pi\hbar}\int_\mu^\infty \frac{\hbar\omega}{4\epsilon^2-(\hbar\omega)^2} d\epsilon .
\end{equation}
This integral has a singularity for $2\hbar\omega>\mu$. Using the residue
theorem, we get
\begin{equation} \label{eq:B_Inter1_final}
\begin{split}
\sigma^{(2)}(\omega,\phi)=&\cos^2(2\phi)\sigma_0 \times \\
&\left[\Theta(\hbar\omega-2\mu)-\frac{i}{\pi}\ln \left|\frac{\hbar\omega+2\mu}{\hbar\omega-2\mu} \right| \right],
\end{split}
\end{equation}
where $\Theta$ is the Heaviside step function. It can be verified that, for
$\phi=0$, Eq.~\eqref{eq:B_Inter1_final} reduces to the formula for graphene.
For $\phi=\pi/4$, the integral is zero, indicating that for the pseudospin-1
lattice, the cone-to-cone transition has no contribution to the optical
conductivity.

\paragraph*{Flat-band-to-cone transition.}
The derivation of the contribution to the optical conductivity by the
flat-band-to-cone transition is similar to that with the cone-to-cone
transition. In particular, for the flat-band-to-cone transition, we have
$E_n=0$ and $E_m=\epsilon$, so
\begin{equation}
\sigma^{(3)}(\omega,\phi)=\sin^2(2\phi)\frac{e^2}{i\pi\hbar}\int_\mu^\infty \frac{\hbar\omega}{\epsilon^2-(\hbar\omega)^2} d\epsilon.
\end{equation}
The singularity now occurs at $\hbar\omega=\epsilon$ with the weight
$\sin^2(2\phi)$. Evaluating this integral, we get
\begin{equation} \label{eq:B_Inter2_final}
\begin{split}
\sigma^{(3)}(\omega,\phi)=&2\sin^2(2\phi)\sigma_0\times \\
&\left[\Theta(\hbar\omega-\mu)-\frac{i}{\pi}\ln \left|\frac{\hbar\omega+\mu}{\hbar\omega-\mu} \right| \right].
\end{split}
\end{equation}

\subsection{Derivation based on Kramers-Kronig formula} \label{Appendix_B2}

As proposed in Ref.~\cite{stauber2013optical}, a different approach to
deriving the optical conductivity is to calculate the real part first and
then use the Kramers-Kronig formula to obtain the imaginary part.
We express the conductivity as the sum of a ``normal'' term and a term
containing a $\delta$-singularity
\begin{equation} \label{eq:B_tot}
\sigma(\omega,\phi)_\text{tot}= \sigma(\omega,\phi) + \pi D\delta(\hbar\omega),
\end{equation}
where $D$ is the Drude weight (or charge stiffness, to be defined below).
To obtain the real part of the ``normal'' term, we use
\begin{equation} \label{eq:B_Fermi}
\begin{split}
&\text{Re} [\sigma(\omega,\phi)]=\frac{1}{2\pi\omega}\iint dk_xdk_y\sum_{n,m}\\
 &[F(E_m)-F(E_n)]j_{nm}^2\delta[\hbar\omega-(E_n-E_m)].
\end{split}
\end{equation}
First, we consider the cone-to-cone transition where $E_n=\epsilon$,
$E_m=-\epsilon$, and the band degeneracy is two. We have
\begin{equation}
\begin{split}
\text{Re} [\sigma^{(2)}&(\omega,\phi)]=\cos^2(2\phi)\frac{e^2}{\pi\omega\hbar^2}
\int_0^{2\pi}\cos^2 \theta_\mathbf{k}  d\theta_\mathbf{k}\times\\
&\int_0^\infty \epsilon [F(-\epsilon)-F(\epsilon)]\delta(\hbar\omega-2\epsilon) d\epsilon.
\end{split}
\end{equation}
The difference in the Fermi-Dirac functions is one only for $\epsilon>\mu$,
and the $\delta$ function is contained in the integration region for
$2\hbar\omega>\mu$. We thus have a step transition at $\mu=2\hbar\omega$:
\begin{equation} \label{eq:B_Inter1_real}
\text{Re} [\sigma^{(2)}(\omega,\phi)]=\cos^2(2\phi)\sigma_0\Theta(\hbar\omega-2\mu).
\end{equation}
Similarly, we obtain, for the flat-band-to-cone transition, the real part of
the conductivity:
\begin{equation} \label{eq:B_Inter2_real}
\begin{split}
\text{Re} [\sigma^{(3)}(\omega,\phi)] &=\sin^2(2\phi)\frac{e^2}{2\pi\omega\hbar^2}\int_0^{2\pi}\sin^2\theta_\mathbf{k} \times \\
& \int_0^\infty \epsilon [f(0)-f(\epsilon)]\delta(\hbar\omega-\epsilon) d\epsilon\\
&=2 \sin^2(2\phi)\sigma_0\Theta(\hbar\omega-\mu),
\end{split}
\end{equation}
which is consistent with the result in Ref.~\cite{illes2015hall}.

To calculate the imaginary part of the conductivity, we use the
Kramers–Kronig (KK) formula~\cite{economou2006green}, which connects the real
and imaginary parts of the response function. There is an additional term in
the integral, which is determined by a cutoff in the case of twisted bilayer
graphene~\cite{stauber2013optical}. The method consists of three steps.

\emph{Step 1}: Calculate the maximum of Re $(\sigma)$ for
$\omega\rightarrow\infty$:
\begin{equation}
\sigma_m(\phi)=\lim_{\omega\rightarrow\infty} \text{Re} [\sigma(\omega,\phi)].
\end{equation}

\emph{Step 2}: Define the Drude weight (or charge stiffness) as
\begin{equation}
D=\lim_{\Lambda\rightarrow\infty}\frac{2}{\pi}\left(\sigma_m(\phi)\Lambda-\int_0^\Lambda \text{Re} [\sigma(\nu,\phi)] d\nu \right).
\end{equation}

\emph{Step 3}: Use the Kramers-Kronig relation to write the imaginary part of
the conductivity as
\begin{equation}
\begin{split}
\text{Im} [\sigma(\omega, \phi)]=&\frac{D}{\hbar\omega}+\\
&\frac{2\hbar\omega}{\pi}\mathcal{P}\int_0^\infty \frac{\text{Re}[\sigma(\nu, \phi)]-\sigma_m(\phi)}{(\hbar\omega)^2-\nu^2}d\nu.
\end{split}
\end{equation}

For the $\alpha$-$\mathcal{T}_3$ lattice, we have
\begin{displaymath}
\sigma_m(\phi)=\sigma_0[2\sin^2(2\phi)+\cos^2(2\phi)],
\end{displaymath}
so the Drude weight is
\begin{equation}
D=\frac{4\sigma_0\mu}{\pi},
\end{equation}
and the Kramers-Kronig relation becomes
\begin{equation} \label{eq:B_imag}
\begin{split}
\text{Im}[\sigma(\omega,\phi)]=&\frac{4\sigma_0\mu}{\pi\hbar\omega}+\frac{2\hbar\omega}{\pi}\mathcal{P}\int_0^\infty \frac{\text{Re}[\sigma(\nu,\phi)]-\sigma_m(\phi)}{(\hbar\omega)^2-\nu^2}d\nu \\
=& \frac{4\sigma_0\mu}{\pi\hbar\omega}-\cos^2 (2\phi) \frac{\sigma_0}{\pi}\ln\left| \frac{\hbar\omega+2\mu}{\hbar\omega-2\mu}\right| - \\
&\sin^2(2\phi)\frac{2\sigma_0}{\pi}\ln\left|\frac{\hbar\omega+\mu}{\hbar\omega-\mu} \right|.
\end{split}
\end{equation}
Together with the real part in Eqs.~\eqref{eq:B_Inter1_real} and
\eqref{eq:B_Inter2_real} as well as the imaginary part Eq.~\eqref{eq:B_imag},
we obtain the same conductivity formulas as these derived based on the
Kubo formula [Eqs.~\eqref{eq:B_Intra_final}, \eqref{eq:B_Inter1_final}, and
\eqref{eq:B_Inter2_final}].

\subsection{Effects of finite temperatures on optical conductivity in the
$\alpha$-$\mathcal{T}_3$ lattice} \label{Appendix_B3}

Consider graphene at a finite temperature, where the Fermi-Dirac distribution
can no longer be treated as a step function. Previous
work~\cite{falkovsky2008optical} gives
\begin{equation} \label{eq:B_temperature1}
\begin{split}
\Theta(\hbar\omega-2\mu) \rightarrow& \frac{1}{2} + \frac{1}{\pi}\arctan\left[\frac{\hbar\omega-2\mu}{2k_BT} \right], \\
|\hbar\omega-2\mu|\rightarrow&\sqrt{(\hbar\omega-2\mu)^2+2(k_BT)^2}. \\
\end{split}
\end{equation}
Comparing Eqs.~\eqref{eq:B_Inter1_final} and \eqref{eq:B_Inter2_final} as well
as the Fermi-Dirac distribution, we can eliminate the factor of $2$ in
Eq.~\eqref{eq:B_temperature1} and study the effects of finite temperature
on the flat-band-to-cone transition through the transformations:
\begin{equation} \label{eq:B_temperature2}
\begin{split}
\Theta(\hbar\omega-\mu) \rightarrow& \frac{1}{2} + \frac{1}{\pi}\arctan\left[\frac{\hbar\omega-\mu}{k_BT} \right], \\
|\hbar\omega-\mu|\rightarrow&\sqrt{(\hbar\omega-\mu)^2+(k_BT)^2}. \\
\end{split}
\end{equation}

Substituting Eqs.~(\ref{eq:B_temperature1}-\ref{eq:B_temperature2}) into
Eqs.~(\ref{eq:2_Intra_final}-\ref{eq:2_Inter2_final}) gives the complete
optical conductivity of $\alpha$-$\mathcal{T}_3$ lattice at a finite
temperature.

\subsection{Effects of impurity scattering on optical conductivity in the
$\alpha$-$\mathcal{T}_3$ lattice} \label{Appendix_B4}

When impurity scattering occurs, a Drude peak will arise in the intra-band
component of the conductivity. For graphene, a previous
work~\cite{gusynin2006unusual} revealed that the Drude peak gives a singularity
at $\omega \rightarrow 0$ of the form $\mu\delta(\hbar\omega)$. Here we derive
the corresponding formula for $\alpha$-$\mathcal{T}_3$ lattice.

Considering the intraband conductivity as given by Eq.~(4) and making the 
change $\omega\rightarrow \omega+i\tau^{-1}$ with $\tau$ being the relaxation 
time, we get
\begin{equation} \label{eq:sigma_tau}
\sigma^{(1)}(\omega,\phi)=\frac{4i\mu\sigma_0}{\pi \hbar (\omega+i\tau^{-1})}.
\end{equation}
Introducing $\Gamma=\hbar\tau^{-1}$, we have
\begin{equation}
\sigma^{(1)}(\omega,\phi)=\frac{4\mu\sigma_0}{\pi}\frac{i(\hbar\omega-i\Gamma)}{(\hbar\omega)^2+\Gamma^2}.
\end{equation}
Taking two successive limits: first $\hbar\omega\rightarrow 0$ and then $\Gamma\rightarrow 0$, leads to the Drude peak
\begin{equation}
\sigma^{(1)}(\omega,\phi))=4\mu\sigma_0\delta(\hbar\omega).
\end{equation}
Together with Eq.~\eqref{eq:B_Intra_final}, we get Eq.~\eqref{eq:2_Intra_final}
in the main text. This result can be verified by inserting the Drude weight
into Eq.~\eqref{eq:B_tot}.

The above derivation can be justified, as follows. First, the result depends 
strongly on the order of limit taking. Note that the Drude singularity is 
represented by a $\delta$-function with a coefficient proportional to the 
chemical potential $\mu$. However, the function is not properly defined for 
$\mu\rightarrow 0$. For graphene in the continuum limit, we have 
$\omega\rightarrow 0$ and $\mu\rightarrow 0$, where the conductivity reaches 
minimum at the Dirac point but the minimal value remains 
unresolved~\cite{sarma2011electronic}. Taking the two limits in the opposite 
order will generate a different result. Nevertheless, for optical waves in 
graphene, because of their high frequency, a positive Fermi energy is required.

Second, the Drude singularity does not depend on $\alpha$, which is consistent 
with the result in Ref.~\cite{tabert2016optical}. That is, for different values
of $\alpha$, the intraband conductivity is invariant. Note that, however, the 
experimental value of the relaxation time for the $\alpha$-$\mathcal{T}_3$ 
lattice is currently unavailable.

Third, for finite impurity scattering, the following substitution holds:
\begin{equation}
\delta(\hbar\omega)\rightarrow \frac{\Gamma^2}{(\hbar\omega)^2+\Gamma^2}.
\end{equation}
About the choice of $\tau$, in Ref.~\cite{jablan2009plasmonics}, the authors 
suggested $\tau=6.4\times 10^{-13}$s. The corresponding conductivity is 
plotted in Figs.~\ref{fig:5}(a) and \ref{fig:5}(b), which reveal that the 
effect of the impurity on the conductivity arises only at low frequencies. 
Heuristically, this is because, in graphene, $\tau$ is of the magnitude 
$10^{-13}$s (corresponding approximately to $10$THz), but optical processes 
in graphene typically occur in the frequency regime above this value. As a 
result, taking into account impurity scattering will not change our result 
appreciably.

We note that an alternative approach to obtaining the Drude singularity was 
provided in Ref.~\cite{stauber2013optical}, in which the authors separated 
the conductivity into two components: a regular component and a term containing
a $\delta$-singularity:
\begin{equation}
\sigma(\omega,\phi)_\text{tot}=\sigma(\omega,\phi) + \pi D\delta(\hbar\omega),
\end{equation}
where $\sigma$ has been derived in Appendix~B2 and $D$ is the Drude weight. 
Substituting the formula of $D$ derived in Appendix~B2 into the equation 
leads to the same result.

\subsection{Imaginary part of the optical conductivity in the low
frequency regime and the effects of impurity scattering} \label{Appendix_B5}

We study the optical conductivity of the $\alpha$-$\mathcal{T}_3$ lattice in 
the small frequency regime. To this end, an earlier work considered the 
imaginary part of the optical conductivity in superconducting materials in the 
zero frequency limit~\cite{marsiglio1996imaginary,jiang1996imaginary}, where 
the results were presented in terms of the product of the frequency and the
imaginary part of the conductivity, which is related to the inverse square
of the penetrate 
depth~\cite{marsiglio1996imaginary,jiang1996imaginary,pronin2001optical} as
\begin{displaymath}
\lim_{\omega \rightarrow 0} \omega {\rm Im} \sigma \propto 1/\lambda_L^2.
\end{displaymath}
The results indicated that, for a low impurity scattering rate, the quantity 
$\omega$Im$\sigma$ can maintain a finite value for a larger frequency interval
near zero. As the impurity scattering rate decreases, the imaginary part of 
the conductivity diverges. Physically, this means that the superconducting 
materials could have a divergent imaginary part of the conductivity. On the 
contrary, for normal states, the quantity $\omega$Im$\sigma$ decays quickly 
to zero and thus converges.

\begin{figure} [ht!]
\centering
\includegraphics[width=\linewidth]{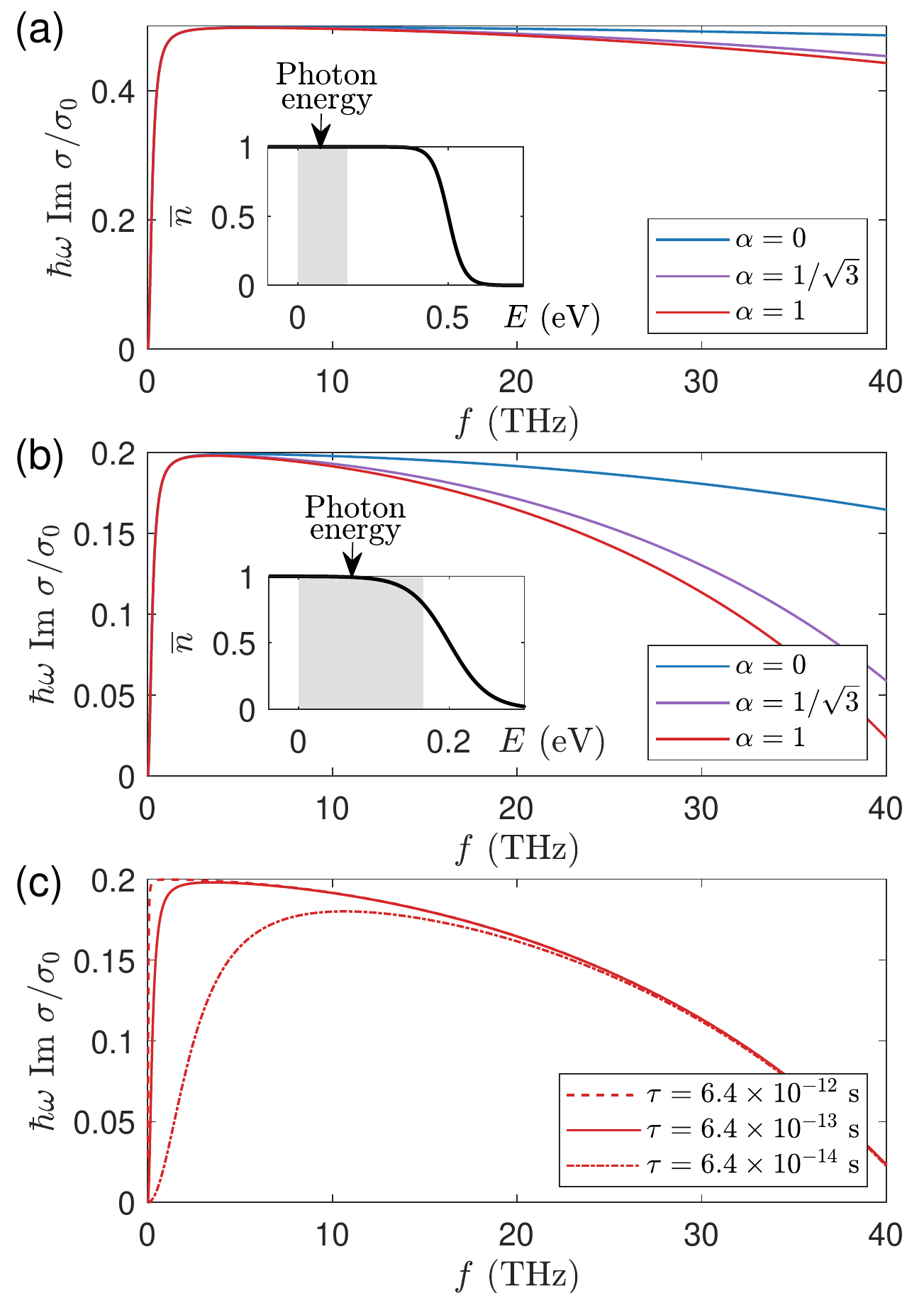}
\caption{Effect of impurity scattering on the optical conductivity
of $\alpha$-$\mathcal{T}_3$ lattices in the low frequency regime. Shown is the 
quantity $\hbar\omega\rm{Im}\sigma/\sigma_0$ (the product between the imaginary
part of the conductivity and the frequency in units of 4eV/$\pi$) versus the 
optical frequency $f$ for different values of the material parameter $\alpha$.  
In (a), the Fermi energy is 0.5 eV. In this case, the range of the photon 
energy (the vertical shaded region in the inset) is far below the Fermi energy
and the flat band states are fully occupied. Interband transitions are unlikely,
so the flat band plays essentially no role in the optical conductivity. In (b), 
the Fermi energy is reduced to $0.2$ eV. In this case, the photon energies as 
indicated by the shaded region in the inset are likely to induce interband 
transitions. The flat band will affect the optical conductivity, as the 
conductivity curves are markedly different for different values of $\alpha$.
The effect of impurity scattering is demonstrated in (c), where the 
conductivity curves for three values of the scattering relaxation time $\tau$ 
are shown for $\alpha=1$. Strong impurity scattering as characterized by a 
relatively small value of $\tau$ can lead to a significant deviation of the 
conductivity curve from the cases of weaker scattering.} 
\label{fig:8}
\end{figure}

For an $\alpha$-$\mathcal{T}_3$ lattice, the two interband transitions as 
described by Eqs.~\eqref{eq:B_Inter1_final} and \eqref{eq:B_Inter2_final} 
vanish at the low frequency limit, since a low energy photon is not able to 
induce an interband transition. For a finite Fermi energy, impurity scattering 
does not change this picture. For the intraband transition in the presence of
an impurity, roughly there are two regimes in Eq.~\eqref{eq:sigma_tau}.
In the first regime, decreasing the frequency $\omega$ will increase the 
driving period but it is still smaller than the relaxation time associated 
with impurity scattering. In this case, it can be seen from 
Eq.~\eqref{eq:sigma_tau} that the product $\omega$Im$\sigma$ converges
to a quantity proportional to the chemical potential. The second regime is
where the frequency decreases further so that the driving period is longer
than the impurity scattering relaxation time. In this case, impurity scattering
dominates and the imaginary part of the optical conductivity approaches zero.

Figure~\ref{fig:8}(a) shows $\hbar\omega\rm{Im}\sigma/\sigma_0$ versus the 
frequency $f$ for $\alpha=0$, $1/\sqrt{3}$ and 1, where the temperature is 
$T=300$ K and the chemical potential is $\mu=0.5$ eV. In this frequency 
range, the corresponding photon energy lies in the interval as indicated by
the shaded region in the inset. We see that the functional behavior of 
$\hbar\omega\rm{Im}\sigma/\sigma_0$ versus $f$ is similar for the three 
different values of $\alpha$. Since the imaginary part of the optical 
conductivity plays an important role in surface wave prorogation and 
scattering, the physical significance is that the flat band has little 
effect on these processes. In this case, the flat band ($E=0$) states are 
full, but any photon energy in the shaded region in Fig.~\ref{fig:8}(a) is
not sufficient to generate an interband transition. Since the flat band 
plays no rule in the optical transitions, different values of $\alpha$ will 
lead to the same optical conductivity.

As the Fermi energy is reduced to, e.g., $0.2$ eV, differences in the 
functional behavior of $\hbar\omega\rm{Im}\sigma/\sigma_0$ versus $f$ begin
to emerge, as shown in Fig.~\ref{fig:8}(b). From the Fermi-Dirac distribution
in the inset, we see that the photon energy is now sufficient to generate 
interband transitions. In this case, the flat band will affect the behavior 
of $\hbar\omega\rm{Im}\sigma/\sigma_0$, but the effect is reduced for small
frequencies. Increasing the temperature will result in a similar effect. 

Figure~\ref{fig:8}(c) shows $\hbar\omega\rm{Im}\sigma/\sigma_0$ versus $f$
for three values of the relaxation time $\tau$ of impurity scattering. For 
relatively large values of $\tau$ (e.g., $\tau = 6\times 10^{-12}$ s or 
$\tau = 6\times 10^{-13}$ s), the quantity $\hbar\omega\rm{Im}\sigma/\sigma_0$
approximately attains the value of the chemical potential in the low frequency
regime before decreasing to zero as the frequency increases. As the value of
$\tau$ decreases (e.g., to $\tau = 6\times 10^{-14}$ s), impurity scattering
begins to dominate, leading to an appreciable change in the functional 
behavior of $\hbar\omega\rm{Im}\sigma/\sigma_0$ versus $f$. We note that 
this effect of impurity scattering on the optical conductivity was previously 
observed~\cite{marsiglio1996imaginary}.

The results in Fig.~\ref{fig:8} can be concisely summarized, as follows.
\begin{itemize}
\item
For fixed impurity scattering rate and Fermi energy, as $\omega$ decreases,
the product $\omega$Im$\sigma$ first reaches some value proportional to the
chemical potential $\mu$ and then attains a different (but convergent) value
after impurity scattering dominates.
\item \vspace*{-0.05in}
For fixed impurity scattering rate and frequency $\omega$, the value of
$\omega$Im$\sigma$ depends on $\alpha$ for a small Fermi energy due to the
enhanced interband transition.
\item \vspace*{-0.05in}
For fixed value of $\alpha$ and Fermi energy, a small impurity scattering rate
(equivalently, a large relaxation time) will lead to divergence of
$\omega$Im$\sigma$. However, for some time $\tau$ that satisfies
$\omega<\tau^{-1}$, the impurity scattering dominates. In this case, the
conductivity is convergent.
\end{itemize}

\section{Scattering cross section from a multilayer sphere} \label{Appendix_C}

To numerically calculate the scattering cross section, we use the iterative
method in Ref.~\cite{johnson1996light}. Here we present the key formulas.

Consider a spherical structure of $L$ layers, where the core is labeled as
$i=1$ and the region outside the structure is denoted as $i=L+1$. In each
layer, the radius is $r_i$ and the refractive index is $m_i$. Define
\begin{equation}
\psi_n(\rho)=\rho j_n(\rho),\text{ } \xi_n(\rho)=\rho h_n^{(1)} (\rho),
\end{equation}
where $j_n$ and $h_n^{(1)}$ are the spherical Bessel functions of the first
and third kind, respectively. Consider the following quantities
\begin{equation*}
\begin{split}
D_n(\rho)&=\frac{\psi_n'(\rho)}{\psi_n(\rho)}, \text{ } G_n(\rho)=\frac{\xi_n'(\rho)}{\xi_n(\rho)}, R_n(\rho)=\frac{\psi_n(\rho)}{\xi_n(\rho)},\\
U_n(r_i)&=m_i\frac{R_n(m_ikr_i) D_n(m_ikr_i)+a_n^{(i)}G_n(m_kkr_i)}{R_n(m_ikr_i)+a_n^{(i)}}, \\
V_n(r_i)&=\frac{1}{m_i}\frac{R_n(m_ikr_i) D_n(m_ikr_i)+b_n^{(i)}G_n(m_kkr_i)}{R_n(m_ikr_i)+b_n^{(i)}}, \\
\end{split}
\end{equation*}
where the coefficients $a_n^{(i)}$ and $b_n^{(i)}$ are to be calculated.
Matching the boundary conditions for each layer leads to the following
iterative equations:
\begin{equation*}
\begin{split}
a_n^{(i+1)}&=-R_n(m_{i+1}kr_i)\frac{U_n(r_i)-m_{i+1}D_n(m_{i+1}kr_i)}{U_n(r_i)-m_{i+1}G_n(m_{i+1}kr_i)},\\
b_n^{(i+1)}&=-R_n(m_{i+1}kr_i)\frac{m_{i+1}V_n(r_i)-D_n(m_{i+1}kr_i)}{m_{i+1}V_n(r_i)-G_n(m_{i+1}kr_i)}.
\end{split}
\end{equation*}
We start from $a_n^{(1)}=b_n^{(1)}=0$ and iterate the above equations until
the final layer is reached. In our computation, the parameter values are
$m_1=\sqrt{\epsilon_1}$ and $r_1=100$ nm for PTFE,
$m_2=\sqrt{\epsilon_{\alpha,N}}$ and $r_2=101$ nm for $\alpha$-$\mathcal{T}_3$
lattice, and $m_3=1$ for free space. The scattering cross section is given by
\begin{equation}
\sigma_\text{sc}=\sum_n\frac{2\pi}{k^2}(2n+1)\left(|a_n^{(3)}|^2+|b_n^{(3)}|^2\right).
\end{equation}


%
\end{document}